\documentclass[sn-basic]{sn-jnl}

\usepackage{appendix}
\usepackage{multirow}
\usepackage{graphicx}
\usepackage{amssymb}
\usepackage{csquotes}
\usepackage[inline]{enumitem}
\usepackage{amsmath}
\usepackage{xcolor}
\usepackage{hyperref}
\usepackage{mathbbol}
\usepackage{booktabs}
\usepackage[usestackEOL]{stackengine}
\newcommand{\myfrac}[3][0pt]{\genfrac{}{}{}{}{\raisebox{#1}{$#2$}}{\raisebox{-#1}{$#3$}}}

\DeclareMathOperator*{\argmax}{arg\,max}
\usepackage{caption}
\usepackage{float}
\usepackage{inputenc}

\begin{document}

\title[ ]{Interactions between the individual and the group level in organizations}
\subtitle{The case of learning and autonomous group adaptation}

\author[1]{\fnm{Dario} \sur{Blanco-Fernandez}}\email{dario.blanco@aau.at}

\author*[2]{\fnm{Stephan} \sur{Leitner}}\email{stephan.leitner@aau.at}

\author[2]{\fnm{Alexandra} \sur{Rausch}
}\email{alexandra.rausch@aau.at}

\affil[1]{\orgdiv{Digital Age Research Centre}, \orgname{University of Klagenfurt}, \orgaddress{\street{Universit\"atsstra{\ss}e 65-67}, \city{Klagenfurt}, \postcode{9020}, \country{Austria}}}

\affil[2]{\orgdiv{Department of Management Control and Strategic Management}, \orgname{University of Klagenfurt}, \orgaddress{\street{Universit\"atsstra{\ss}e 65-67}, \city{Klagenfurt}, \postcode{9020}, \country{Austria}}}

\abstract{Previous research on organizations often focuses on either the individual, team, or organizational level. There is a lack of multidimensional research on emergent phenomena and interactions between the mechanisms at different levels. This paper takes a multifaceted perspective on individual learning and autonomous group formation and adaptation. To analyze interactions between the two levels, we introduce an agent-based model that captures an organization with a population of heterogeneous agents who learn and are limited in their rationality. To solve a task, agents form a group that can be adapted from time to time. We explore organizations that promote learning and group adaptation either simultaneously or sequentially and analyze the interactions between the activities and the effects on performance. We observe underproportional interactions when tasks are interdependent and show that pushing learning and group adaptation too far might backfire and decrease performance significantly.}


\maketitle

\section{Introduction}\label{sec:introduction}

Learning in organizational contexts and issues related to dynamic capabilities are usually researched at different (and often isolated) levels. First, at the level of the individual, research mainly addresses the enablers of learning and adaptation, contingencies, and the effects of learning and adaptation for the individual \citep{Jyothibabu2010, Murray2005}. Second, at the organizational level, research is often concerned with developing and maintaining learning systems, institutionalizing learning in terms of embedding it into the processes, structures and strategies effective in the organization, promoting learning, and issues related to how organizational capabilities emerge \citep{Fiol1985,Teece1997}. Third, at the intermediate level, the importance of groups to link the individual and the organizational level is usually emphasized. It is addressed how individually learned information can be integrated, shared and adjusted \citep{Murray2005}. Research on learning at the intermediate level is, for example, concerned with how to compose groups ideally, how to use individual knowledge optimally, and how to take advantage of synergies \citep{Bell2017,Licalzi2012}.
However, since the capabilities at the individual level might be dynamic because of learning, the composition of groups at the intermediate level may have to be dynamic too. Thus, the layers are apparently interrelated. 
Previous studies often take an unidimensional perspective and focus on one level only. This lack of integration across levels reflects the micro-macro divide that is predominant, for example, in the field of managerial science \citep{Aguinis2011,Molloy2011}. The focus on only one of the interrelated levels can, to some extent, be explained by disciplinary borders \citep{Bargiela-Chiappini2002}. Research in psychology, for example, tends to focus on issues related to learning at the individual level. In contrast, research on organizational design is likely to be more interested in matters concerning the collective level. To overcome the gap between micro- and macro-level research, \cite{Hitt2007} recommend \textit{(i)} applying a multi-level design to existing models, \textit{(ii)} considering the consequences of micro-level activities for macro-level performance, \textit{(iii)} pushing disciplinary boundaries, and \textit{(iv)} addressing problems of practical relevance.  

Our research follows the suggestions provided by \citet{Hitt2007}. We apply a multi-level approach that connects the individual and group level. In particular, we investigate whether individual learning and adaptive group composition mutually reinforce or attenuate each other concerning their effects on performance. We aim at answering the following research questions: 
\textit{(i)} How do individual learning and group adaptation interact and affect task performance if they are promoted \textit{simultaneously}?
\textit{(ii)} If individual learning and group adaptation are promoted \textit{sequentially}, what are the effects on performance in the sequential stages?
\textit{(iii)} Are there any moderating effects of task decomposability, i.e., the fact that the tasks assigned to different agents are interdependent or not? 

To answer these questions, we propose an agent-based model with a population of heterogeneous agents who are limited in their rationality \citep{Simon1957}. In our model, the limitations imposed on agents are two-fold: First, the agents' abilities to solve tasks are restricted to a particular area of expertise. In an organizational context, we consider agents who are experts in fields such as accounting, marketing, or production management. Second, agents have limited information within their area of expertise and limited cognitive capabilities. Consequently, they cannot oversee the entire room of actions immediately but explore it only sequentially. This means they \textit{learn} new ways to perform their tasks over time \citep{Leitner2020}. The population of agents autonomously forms a group following an auction-based mechanism, and agents can re-organize their groups from time to time. We base our model on the well-known $N\!K$-framework that allows placing the agents in task environments of different complexity \citep{Levinthal1997,Wall2020}. We deem our multi-level approach to analyze interactions between individual learning and group adaptation a relevant contribution to bridging the gap between the micro- and macro-level in managerial science. Our results are also of relevance for decision  making in organizations since such multi-level approaches appear to be particularly interesting for corporate practice \citep{Aguinis2011}.    

The remainder of this paper is structured as follows: We discuss the related literature on dynamic capabilities and multi-level considerations, individual learning and bounded rationality, and group formation and adaptation in Sec. \ref{sec:background}. The simulation model is introduced in Sec. \ref{sec:model}. The results of the simulation experiments are presented in Sec. \ref{sec:results} and discussed in Sec. \ref{sec:discussion}. Finally, Sec. \ref{sec:conclusion} concludes the paper. 

\section{Related research}\label{sec:background}

\subsection{Dynamic capabilities and multi-level considerations}
\label{sec:dyn-capa}

An organization's dynamic capabilities are usually regarded as some sort of higher-order capability that affects how tasks are solved \citep{Winter2003}. Dynamic capabilities are manifold, making it particularly difficult to find a coherent and tangible definition \citep{Spanuth2020}. We follow the conceptualization provided in \cite{Zollo2002}, according to which dynamic capabilities are the stable and learned patterns of activity within an organization, through which an organization modifies its routines. In the dynamic-capabilities framework, the learning of the routines needs to take place at both the individual and organizational level and is driven by a steady demand for adaptation to the task environment. Naturally, having learned the appropriate patterns of activity substantially increases task performance \citep{Eisenhardt2000,Teece1997}. This means that individuals and organizations need to reconfigure their capabilities (e.g., knowledge, skills, abilities) to meet the requirements for solving the tasks they face efficiently and successfully. \citet{Bendig2018} highlight that there are two parallel developments in this research context: The first stream of research focuses on the micro-foundations of how capabilities evolve and aims to understand how learning by individual employees aggregates to an organization's capability \citep[see, e.g., also][]{Abell2008}. The second stream of research exclusively focuses on the macro-level and analyzes how managerial decisions at the top level affect an organization's performance \citep{Helfat2015}. While both streams substantially advanced the knowledge on dynamic capabilities, the interaction between the two levels is still under-researched \citep{Bendig2018,Fainshmidt2017}. 

Research related to the emergence of dynamic capabilities is tightly connected to the literature on learning in organizational contexts. The latter distinguishes between three levels of learning, i.e., \textit{(i)} individual learning, \textit{(ii)} team learning, and \textit{(iii)} organizational learning \citep[see, e.g., ][]{Edmondson2001,Kirkman2000,Murray2005,Senaratne2011}. By doing so, the literature on learning adds an intermediate layer to the two levels already considered in the context of dynamic capabilities, i.e., a group level between the individual and the organizational level. \citet{Murray2005} emphasize the role of the intermediate layer and particularly underline the role of learning at the level of teams as a link between individual and organizational learning. Also, it is widely agreed in previous research that individual learning is the foundation for learning at higher levels and that the learning outcomes at the collective level are more than the accumulation of individual learning \citep{Casey2005,Dodgson1993,Garratt1987,Hedberg1981,Miller1996,Popper2000}.\footnote{For reviews of the literature on learning in organizational contexts, the reader is referred to \cite{Bapuji2004}, \cite{Basten2018}, \cite{EasterbySmith2000}, and \cite{Odor2018}, amongst others. Reviews related to the dynamic capabilities framework are, for example, provided by \citet{Barreto2010}, \citet{Schilke2018}, and \citet{Wang2007}.}  Within this three-layer framework, our focus follows an argument put forward by \citet{Simon1991}. He claims that learning in organizations may occur in two ways: First, the individual members of an organization may learn. Second, the organization may ingest new members who have new knowledge that was not available to the organization before. In line with this argumentation, we focus on understanding the effects of a variation in two organizational design parameters, namely \textit{(i)} individual learning and \textit{(ii)} group adaptation in the sense of changing a group's composition from time to time. For this purpose, we consider the probability of learning at the individual level and groups of different lifetimes. Our research analyzes how efficiently agents use the information they have previously learned and how promoting learning and group adaptation affects the performance of organizations.

\subsection{Individual learning and bounded rationality}
\label{sec:background-learning}

In the literature on learning in organizations, there are two main streams of research. The first stream focuses on the enablers of learning, the second on the results of learning in organizations \citep{Jyothibabu2010}. Research focusing on the enablers of learning explores and discusses ways and means to promote individual learning. Such ways and means comprise, for example, the methods of mentoring \citep{Lankau2007}, learning by doing and/or exploration \citep{Arrow1971,Beugelsdijk2008}, fostering a creative working environment \citep{Annosi2020,Oldham2003}, employee training \citep{Salas1999,Tharenou2007}, and the design of information flows \citep{Cohen1991}. In this paper, we are not concerned with the efficiency of the different ways and means to promote learning but rather consider them as given. We locate our research in the second stream. Hence, we particularly emphasize the effects of promoting individual learning on performance, its interplay with promoting group adaptation, and the results for an organization. Thereby, we contribute to closing the gap between (changes in) micro-level behavior, i.e., the behavior of individual agents, and the organization's macro-level outcome \citep{Aguinis2011}. 

Following the concept of dynamic capabilities, individual learning allows for adapting to the requirements posed on the organization by dynamic environments, which, in turn, is key to organizational competitiveness and survival \citep{Teece1997}. As soon as we break down the problem of adaptation to smaller entities within the organization -- such as groups or individual decision makers -- and consider decentralized decision-making authority, a similar but extended argumentation applies: From the perspective of the individual, it is not just important to adapt to the organizational environment by learning, but the individual agent is also well-advised to adapt to their individual environment when making decisions. This individual environment captures, for example, the decisions made by fellow agents within a group. This is particularly relevant if the tasks assigned to different decision makers are interdependent \citep{Rivkin2007}.  

Learning at the individual level is affected by characteristics such as cognitive capability, learning styles, and interpretative ability \citep{Crossan1999,Murray2005,Neisser2014}. Research often considers limitations in these characteristics and addresses, amongst others, decision makers who over-weight information in favor of prior beliefs \citep{Darley1983}, slant information towards a preferred state \citep{Kunda1990}, and over- or under-react to information \citep{Leitner2017}. More generally speaking, research often assumes that individual agents suffer from the limitations of bounded rationality in the sense of \citet{Simon1957}, to which \citet{Hendry2002} refers as \enquote*{incompetence}. In particular, he argues that humans might have limited knowledge and foresight and face limitations of rational understanding and communication that may arise from language, culture, and cognition \citep[see also][]{Martin1993}. Of course, this directly translates into consequences for individual learning. First, suppose humans suffer from limited foresight. In this case, they might have problems in correctly predicting the outcomes of their future actions, or they might not be able to form beliefs about the actions of others \citep{Enke2019}. Second, if the cognitive capabilities are limited, decision makers might not be able to understand and oversee their entire room for actions. In practical terms, one might not be aware of all feasible ways to carry out a specific task at a time but rather sequentially explore this room \citep{Leitner2020}. Third, the extent to which learning is successful might be affected, amongst others, by the individuals' cultural backgrounds \citep{Kim2014} or their technical literacy \citep{Qureshi2009}. 

There is a long tradition of studying (individual) learning in organizational contexts. \cite{March1991}, for example, relates decision making to a process by which agents balance learning new solutions to a particular task (i.e., \textit{exploration}) and building on the solutions they already know (i.e., \textit{exploitation}). By choosing a specific mix of both strategies, agents adapt more or less successfully to the organizational and individual environments. Previous research suggests that an appropriate balance of exploration and exploitation is the key to improving task performance \citep{Levinthal1997,March1991,Rivkin2003}. In particular, it has been found that exploration is important for increasing task performance in the early stages of task-solving, but exploitation becomes more relevant in later periods \citep{Leitner2021a,Levinthal1997,Yang2007}.

\subsection{Group formation and adaptation}
\label{sec:background-adaptation}

In Sec. \ref{sec:dyn-capa} we addressed the intermediate level of the group that plays a pivotal role in organizational learning since it links the individual and the organizational level. One fundamental question in this context is how groups should be ideally composed \citep{Higgs2005}. \citet{Mello2006} argue that heterogeneous groups perform better than homogeneous groups since heterogeneity assures access to various information. The argument that heterogeneity might increase the performance is in line with the findings presented in \cite{Licalzi2012}, who analyze the power of heterogeneous agents joining forces in a group to solve tasks over a large solution space. They find, for example, that larger groups can solve problems that individual agents cannot solve alone. They also claim that positively correlated abilities of agents require larger groups to solve tasks efficiently. However, they also show that teaming up does not necessarily guarantee success in all cases. Further empirical evidence corroborates this finding by relating the poor performance of heterogeneous groups to differences in the agents' abilities and characteristics and to heterogeneous preferences, which cause different behaviors and objectives \citep{Ancona1992,Bertrand2003}.

The characteristics that make a group heterogeneous are manifold. Previous research argues that heterogeneity might result from differences in the social background, age, gender, education, national culture, and professional development, amongst others \citep{Bell2007,Bell2017,Hoffman1961,Hofstede1991,Mello2006}. \citet{Krech1962} provide a systematic analysis of the variables that might affect the performance of groups and clusters them into four categories: \textit{(i)} structural variables (characters, talent, size, etc.), \textit{(ii)} situative environmental variables (e.g., functional position), \textit{(iii)} task-related variables (type of task, restrictions in, e.g., time, etc.), and \textit{(iv)} intervening variables (personal relations, level of interaction, etc.). 
Following the categorization provided in \cite{Krech1962}, we focus on the role of the \textit{(i)} structural variables in group heterogeneity. Also in line with the arguments brought forward in \cite{Mello2006}, we consider the professional background by explicitly modeling agents that are experts in specific fields, such as accounting, marketing, or production management.
We do not take into account the \textit{(ii)} situative and \textit{(iv)} intervening variables. However, our research actively controls for the \textit{(iii)} task-related variables by modeling tasks of different degrees of decomposability and complexity.

We follow the argument raised by \citet{Simon1991} and use the intermediate group level to endow the group with knowledge that was not available earlier. We do so by ensuring that only those agents join forces in a group who are best prepared for the task they face. Groups can be formed either by a top-down or a bottom-up approach. The top-down approach corresponds to the idea of classical organizational design and considers that managers conceptually develop a group's composition before implementation \citep{Romme2003,Simon2019}. The bottom-up approach, on the contrary, follows an evolutionary perspective and regards a group's composition as an emergent property \citep{Tsoukas1993}. This corresponds to the ideas of plastic control \citep{Popper1978} and guided self-organization \citep{Prokopenko2009}. In our paper, we follow the bottom-up approach to group formation and implement a corresponding mechanism based on a second-price auction. The design of this mechanism is inspired by previous research in the fields of robotics and transportation research, which also employs auction-based mechanisms for bottom-up task allocation and collaboration \citep{Dai2011,Ng2020,Rizk2019}. Since we consider agents with dynamic capabilities, a group composition yielding the best possible results at one particular point in time is not necessarily the optimal composition at another point in time. This argumentation is in line with research on the interface between dynamic capabilities and temporary organizations. \citet{Spanuth2020}, for example, claim that \textit{temporary} structures enhance an organization's innovative capacity and strategic flexibility, finally resulting in better performance. We account for this relationship by controlling for a group's lifetime and analyzing the effects of different lifetimes on performance. Examples of autonomously formed (temporal) groups can, amongst others, be found in the context of agricultural cooperatives \citep{Hannachi2020}, consulting firms \citep{Creplet2001}, and professional services partnerships \citep{Gershkov2009}.

\section{The Model}\label{sec:model}

We set up an agent-based model of an organization formed by a population of $P=30$ agents. This organization aims at solving a particular task. We endow the agents with (heterogeneous) capabilities related to specific areas of expertise.\footnote{We have implemented the agent-based model in Python 3.7.4.} 
The agents are limited in their cognitive capacities. For example, they might have limited cognitive resources and they suffer from restrictions in information processing. As a consequence, they cannot handle the task alone but have to collaborate with other agents. That is why a subset of the agent population autonomously forms a group of $M \in \mathbb{N}$ members who jointly solve the task. The group-formation mechanism follows the idea of a second-price auction. Depending on the studied scenario, we allow for individual learning (see Sec. \ref{sec:agents}) and changing the group's composition from time to time (see Sec. \ref{sec:auction}). We run simulations and observe the agents' behavior and the task performance achieved by the group over $t=\{1,\dots,T\}\in \mathbb{N}$ periods. In the following subsections, we introduce the model's three main building blocks: \begin{enumerate*}[label=\textit{(\roman*)}]
\item the task environment in Sec. \ref{sec:task-environment},
\item agents and individual decision making in Sec. \ref{sec:agents}, and 
\item the group-formation mechanism in Sec. \ref{sec:auction}. 
\end{enumerate*} Section \ref{sec:variables} discusses the key parameters and the sequencing during simulation runs. Finally, in Sec. \ref{sec:observation}, we introduce the performance measures. 

\subsection{Task environment} 
\label{sec:task-environment}

\paragraph {Task and performance contributions} 
We represent the task environment by a performance landscape based on the  $N\!K$ framework with an $N$-dimensional decision-making task and $K$ interdependencies among decisions. We denote the task by the binary vector
\begin{equation}
    \mathbf{d}=\left(d_1, \dots, d_N \right)~,
\end{equation}
consisting of $N$ individual decisions $d_n$, where $n=\{1,\dots,N\} \in \mathbb{N}$ and $d_n \in \left[0,1 \right]$. For this paper, we set $N=12$. Thus, there are $2^{12}$ feasible solutions to $\mathbf{d}$ that all take the form of a 12-digit binary string.\footnote{For readability, we suppress the subscript $t$ in Sec. \ref{sec:task-environment}.} 

There are $K \in \mathbb{N}_0$ interdependencies among the individual decisions that determine task \textit{complexity} and \textit{decomposability} (i.e., the extent to which tasks are decomposable). Higher values of $K$ stand for more interdependencies and, thus, higher task complexity. In our framework, the interdependencies indicate that the outcome of a single task is not only affected by the decision associated with this task but also by $K$ other decisions \citep{Levinthal1997}. The decomposability is affected by the patterns of interdependencies. We denote the performance contribution of an individual decision $d_n$ by the pay-off function
\begin{equation}
    c_n=f(d_n, d_{i_1}, \dots, d_{i_K})~, 
\label{eq:cont}
\end{equation}
\noindent where $\{i_1, \dots, i_K \} \subseteq \{1, \dots, n-1, n+1, \dots, N \}$. Following the $N\!K$ framework, we draw the performance contributions of individual decisions from a uniform distribution so that $c_n \sim U(0,1)$ \citep{Weinberger1989}.

In line with the $N\!K$ framework, we compute the overall task performance as the mean of all individual performance contributions, such that 
\begin{equation}
    C \left( \mathbf{d} \right)= \frac{1}{\| \mathbf{d} \|} \sum^{ \|\mathbf{d} \|}_{n=1} c_{n}~,
    \label{eq:perf}
\end{equation}
\noindent where $\|\mathbf{d}\|$ indicates the length of vector $\mathbf{d}$.

\paragraph{Subtasks} Recall that the population of agents forms a group of $M$ members who jointly perform the task. We symmetrically divide the $N$-dimensional task into $M$ subtasks of size $S=N/M$, and we sequentially assign agents a subtask.\footnote{Please note that we allocate tasks to agents following a bijective function.} Thus, within the group, all agents $m=\{1,\dots,M\}\in\mathbb{N}$ are assigned their subtasks, such that 
\begin{equation}
    \label{eq:subtasks}
    \mathbf{d}_m = \left(d_{S\cdot(m-1)+1},\dots, d_{S\cdot m}  \right)~.
\end{equation}
\noindent With a task consisting of $N=12$ decisions and a group of $M=3$ agents, agent $1$ is responsible for individual decisions $1$ to $4$, agent $2$ is in charge of decisions $5$ to $8$, and decisions $9$ to $12$ are assigned to agent $3$ (see Fig. \ref{fig:group-formation}).
The subtasks can be related to specific \textit{areas of expertise} that reflect the need for different skills required to solve tasks \citep{Hsu2016}. During the phase of model initialization, we randomly place every agent in the population in one area of expertise. Agent $1$ might, for example, be an expert in accounting, while agents $2$ and $3$ could be experts in, e.g., marketing and production management, respectively. Once placed in an area of expertise, the agents' respective capabilities are limited to performing tasks that are within this area \citep{Giannoccaro2019,Hsu2016,Rivkin2003}. 

\paragraph{Task decomposability}

\begin{figure}
 \centering
 \includegraphics[width=0.85\textwidth]{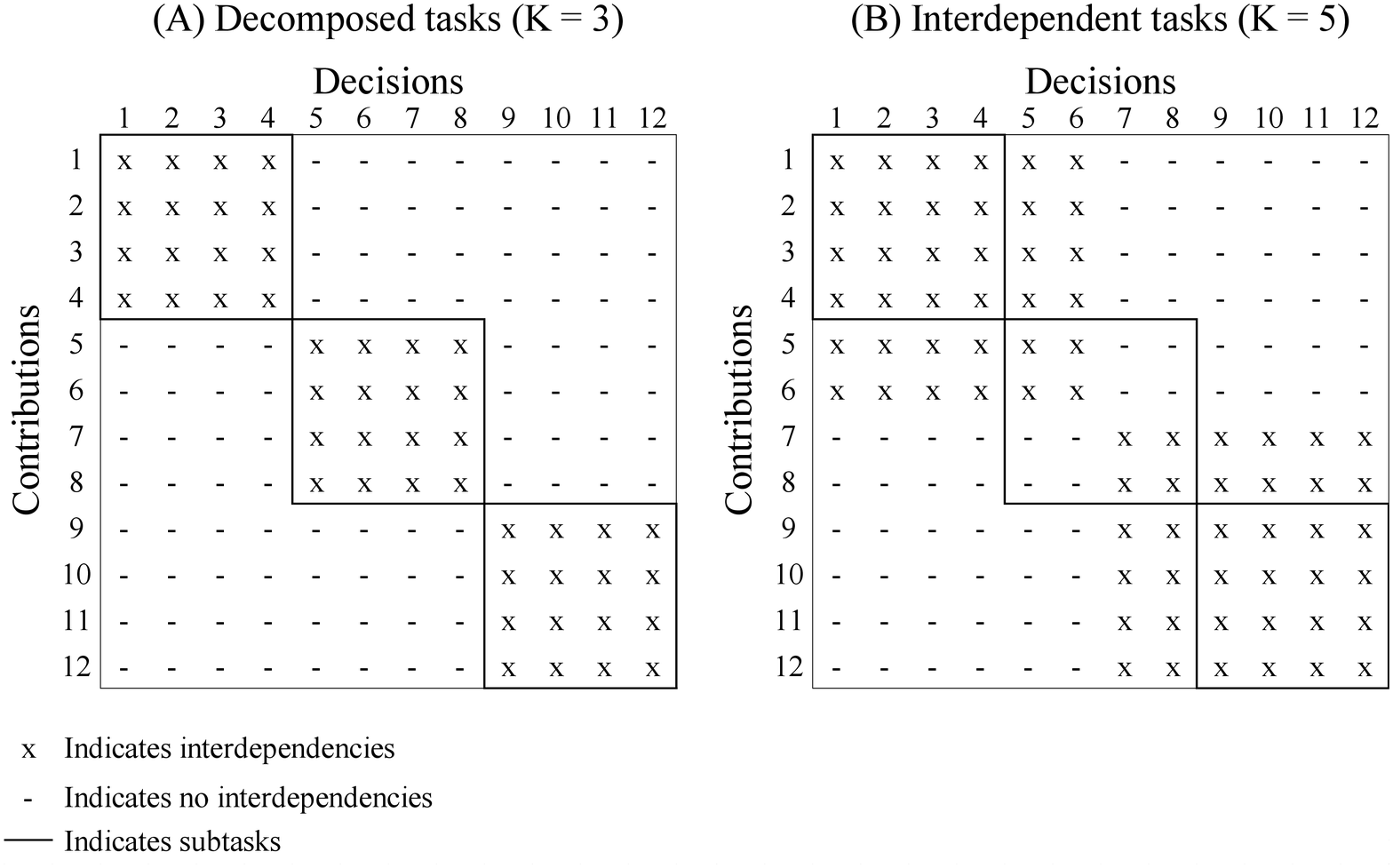}
 \caption{Interdependence matrices.}
 \label{fig:matrix}
\end{figure}

Since our aim is to understand the moderating role of task decomposability, we explicitly control for the patterns of interdependencies. We consider the following two cases: 
\begin{itemize}
    \item In the case of a \textit{decomposable task}, every decision affects the contributions of three other decisions ($K=3$) of one subtask. Figure \ref{fig:matrix}(A) \textit{(Decomposed tasks)} illustrates this case. The solid lines and \enquote*{x} indicate the subtasks and interdependencies, respectively. There is complete interdependence within subtasks but no cross-interdependencies with other agents' subtasks. Thus, one agent's individual decisions do not affect the performance contributions associated with decisions assigned to other agents.
    \item In the case of an \textit{interdependent task}, each decision affects the contributions of five other decisions ($K=5$). Interdependent tasks are characterized by cross-interdependencies between the agents' respective subtasks. This structure implies that an agent's decisions also affect the performance contributions associated with decisions assigned to other agents (see Fig. \ref{fig:matrix}(B), \textit{Interdependent tasks}).
\end{itemize}
\paragraph{Performance landscape} 
The number of decisions $N$, the task complexity $K$, and the performance contributions jointly determine a particular \textit{performance landscape}. We use the pay-off function introduced in function Eq. \ref{eq:cont} to map the $2^N$ feasible solutions to a performance landscape. If there are no interdependencies between individual decisions ($K=0$), the resulting performance landscape has a single peak. Increasing $K$ to its maximum of $K=N-1$, results in maximally complex performance landscapes since altering one single decision would affect the outcomes associated with all tasks. Consequently, if $K$ increases, the performance landscape gets more rugged with numerous local maxima in the extreme case \citep{Altenberg1994,Rivkin2007}. By making decisions, the group moves in the performance landscape and follows the objective to increase its performance step-wise. The following section introduces the agents' characteristics and their decision rules. 

\subsection{Agents and individual decision making} 
\label{sec:agents}

\paragraph{Agents' characteristics}

Recall that we model a population of $P=30$ \textit{heterogeneous} agents \citep{Simon1957}. Bounded rationality imposes limits on them in two respects. First, agents are limited in their cognitive capacity and can only perceive a subtask but not the entire decision problem. In consequence, agents are experts in one area such as accounting, marketing, or production management. Also, agents do not monitor the history of solutions to the decision problem, which results in myopic agents who only optimize their immediate utility and refrain from making forecasts based on the history of solutions \citep{Artinger2021,Leitner2020}. Second, the limitation in the cognitive capacity also affects the agents' search behavior and memory. Our model considers agents who, when making their decisions, cannot oversee the entire solution space but only possess the cognitive capabilities to evaluate some solutions at a time. We endow agents with the ability to explore feasible solutions to their subtasks sequentially and forget already explored solutions because of their limited memory.

\paragraph{Individual learning}

To overcome the limitations at the level of the performance components, agents explore the solution space over time, i.e., they learn.\footnote{Please note that learning could take place at multiple levels \citep{Kim1998}. We exclusively focus on the individual agent.} Recall, Eq. \ref{eq:subtasks} denotes agent $m$'s subtasks by $\mathbf{d}_m$. Since we model binary decisions, the set of feasible solutions to subtask $\mathbf{d}_m$ includes $2^{\|\mathbf{d}_m\|}$ solutions, which we refer to as solution space. Above, we argue that agents cannot oversee the entire solution space at a time. Let us refer to the solutions that agent $m$ is aware of in period $t$ by 
\begin{equation}
\label{eq:solution-space}
{\mathbf{S}}_{mt} =\left(\hat{\mathbf{d}}_{m1},\dots,\hat{\mathbf{d}}_{mI}\right)~,
\end{equation}
where $\hat{\mathbf{d}}_{mi}$ are bitstrings that represent feasible solutions to subtask $\mathbf{d}_m$, $i=\{1,\dots,I\}\in \mathbb{N}$ and $1 \leq I \leq 2^{\|\mathbf{d}_m\|}$. If agent $m$ is, for example, an expert in production management, ${\mathbf{S}}_{mt}$ represents all possible ways to organize the production process she is aware of in period $t$. Consequently, the symbol $I$ is a proxy for agent $m$'s cognitive capacity. The higher (lower) $I$, the more (fewer) solutions the agent knows to the subtask. 
The known solutions are dynamic. At every period, agents might learn new solutions that differ in one bit from any of their already known solutions, or they forget solutions that are not utility-maximizing in the current period. Learning and forgetting are independent events that occur with a fixed probability $\mathbb{P}$. We consider the following three scenarios:

\begin{itemize}
    \item In the case of \textit{zero individual learning}, agents neither learn nor forget any solution they already know, so we set $\mathbb{P}=0$.
    \item In the case of \textit{moderate individual learning}, we set $\mathbb{P}=0.25$.
    \item In the case of \textit{high individual learning}, we set $\mathbb{P}=0.5$. 
\end{itemize}

\noindent Initially, agents are aware of one solution to their subtask. Depending on the value of $\mathbb{P}$, the agents' known solution spaces are more or less dynamic. Increases in the value of $\mathbb{P}$ could indicate that organizations support employees in learning, e.g., they provide learning resources, training, or incentives for the creation of new products, procedures, and methods \citep{Creplet2001}. We refer to the actions taken by the organization to increase the probability of learning (i.e., to increase $\mathbb{P}$) as \textit{promoting individual learning.}

\paragraph{Individual decision-making rule}

In every period, every agent $m$ who is part of the group can decide which solution in her solution space ${\mathbf{S}}_{mt}$ she wants to implement. We denote the solution to agent $m$'s subtask implemented in period $t$ by $\mathbf{d}_{mt}$. We formalize the corresponding decision rule in Eq. \ref{eq:agents-decison}.

Agent $m$'s utility in period $t$ results from the performance contributions generated by the implemented solution to her subtask (i.e., her \textit{own performance}) and the performance generated by the solutions implemented by the remaining agents $r=\{1,\dots,M\}\in \mathbb{N}$, where $r\neq m$. We refer to the latter as \textit{residual performance} and denote the other agents' solutions by
\begin{equation}
\label{eq:residual-decisions}
\mathbf{D}_{mt} = \left(\mathbf{d}_{1t}, \dots, \mathbf{d}_{\{m-1\}t},\\ \mathbf{d}_{\{m+1\}t},\dots, \mathbf{d}_{Mt} \right)~.
\end{equation}
Then, agent $m$'s utility follows the linear function
\begin{equation}
    U \left( \mathbf{d}_{mt}, \mathbf{D}_{mt} \right) = \alpha \cdot C(\mathbf{d}_{mt}) + \beta \cdot \frac{1}{M-1} \sum_{\substack{{r=1}\\{r\neq m}}}^{M} C(\mathbf{d}_{rt}) ~.
\end{equation}
\noindent We compute the performance in line with Eq. \ref{eq:perf}. The agent's utility is affected by a linear incentive scheme that is parameterized by $\alpha \in \mathbb{R}$ and $\beta \in \mathbb{R}$ to weight agent $m$'s own and residual performances, respectively, and $\alpha + \beta = 1$.

Every agent's objective is to maximize their utility, which is only possible by participating in the group.\footnote{Agents who are no group members receive zero utility, and we omit outside options.} We omit the coordination of decisions between agents and allow them to act autonomously. Consequently, agent $m$ is not aware of the solutions that the other $r \neq m$ group members intend to implement in a period. However, agents can observe the solutions implemented in the previous period, $\mathbf{D}_{m\{t-1\}}$. Agents use this information and base their decisions in $t$ on the \textit{estimated utility}, for which it is assumed that the residual decisions do not change from the previous period. Consequently, agent $m$'s decision rule takes the form of
\begin{equation}
\label{eq:agents-decison}
    \mathbf{d}_{mt} := \argmax_{d^\prime \in \mathbf{S}_{mt}} ~ U\left( d^\prime, \mathbf{D}_{m\{t-1\}} \right)~.
\end{equation}
\noindent The function \enquote*{arg max} returns the argument that maximizes the utility function.

\paragraph{Group solution}
Once all agents have made their decisions, we compute the entire solution to the task in period $t$ by
\begin{equation}
\label{eq:groupsolution}
    \mathbf{d}_{t} := \mathbf{d}_{1t}^{} {}^\frown \dots {}^\frown \mathbf{d}_{Mt} ~,
\end{equation}

\noindent where $^\frown$ indicates the concatenation of the solutions individually implemented by the agents. Once the entire solution $\mathbf{d}_t$ is implemented, all agents in the population can observe it.

\subsection{Group-formation mechanism}
\label{sec:auction}

The model considers that the population of agents autonomously forms \textit{one} group consisting of $M$ agents. In a group, there is one agent per area of expertise. Thus, if the task at hand requires experts from accounting, marketing, and production management, a group-formation mechanism makes sure that one experts per area joins the group. 

\begin{figure}
 \centering
 \includegraphics[width=0.8\textwidth]{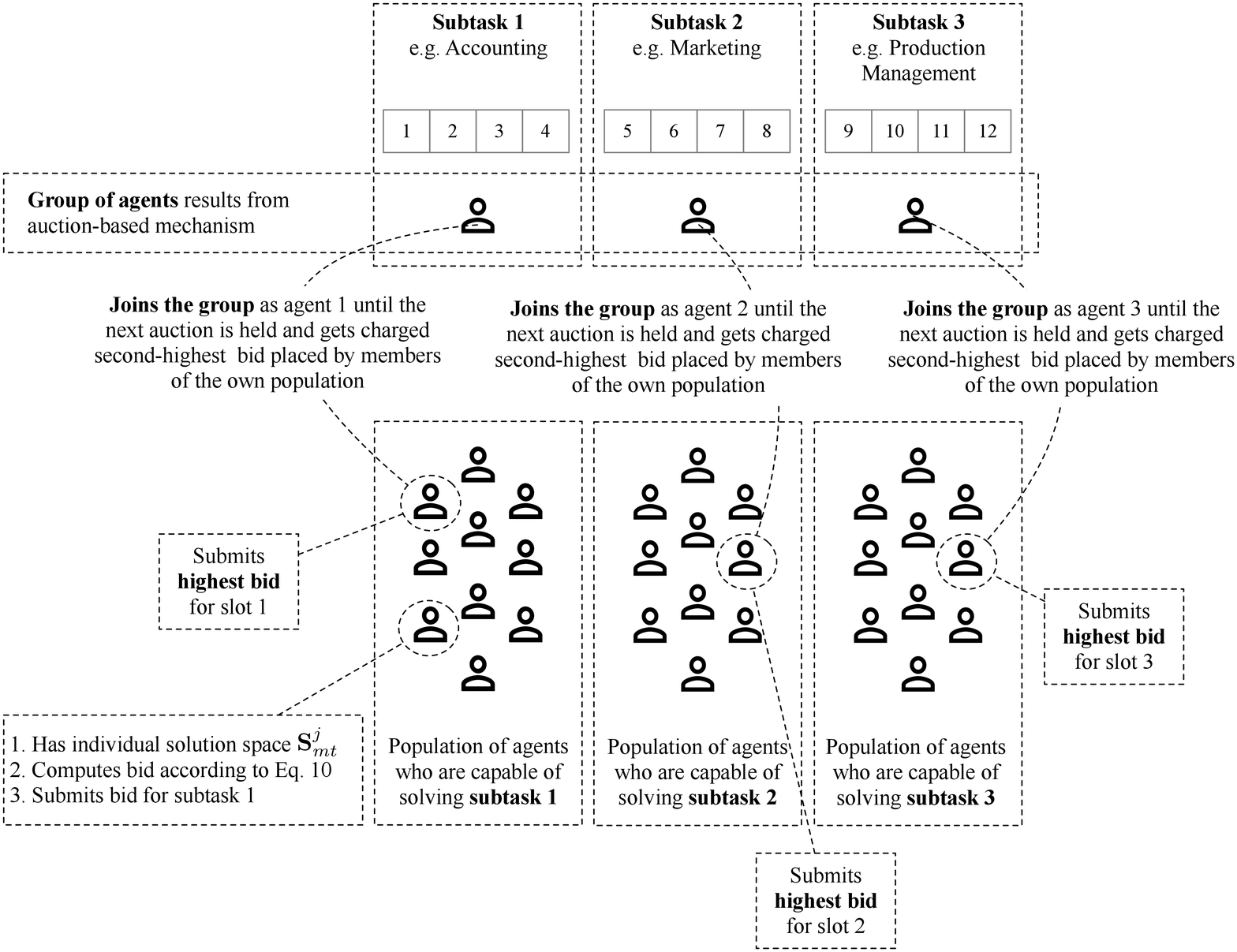}
 \caption{Group formation process within the organization}
 \label{fig:group-formation}
\end{figure}

Recall, we consider an organization formed by $P=30$ agents, and we split the overall task into $M$ subtasks. We symmetrically allocate the agent population to areas of expertise. $P_m=P/M$ potential candidates could solve a particular subtask and, consequently, be a group member.\footnote{Please note that subscript $m=1,\dots,M$, thus, indicates (i) the group members assigned to a subtask, (ii) the subtask, and (iii) the subsets of the population of agents that are capable of solving the subtasks.} Thus, with a population of $30$ agents and three subtasks, there would be ten experts that could solve a particular subtask.
Let us denote the agents who possess the capabilities to solve the subtask $\mathbf{d}_m$ by $p_{m}^{j}$, where $j=\{1,\dots,P_m\}\in \mathbb{N}$. The challenge is to identify those agents in the areas of expertise who are best prepared to solve the task in a group. To do so, agents employ a mechanism that follows the concept of a second-price auction (see also Fig. \ref{fig:group-formation}). Every time an auction is held, the agents use the following procedure to form a group \citep{Rizk2019}:
 \begin{enumerate}
     \item Agents are informed about the auction, and they can place bids to join the group. Since agents can only experience utility by joining the group, every agent has the incentive to participate in the auction. Auctions are anonymous, and agents have no information about the other agents' bids. The bids are independent, and the \enquote*{price} of joining the group is determined by the bids only.
     \item Agents compute their bids by drawing on the information available to them in the following way: In line with Eq. \ref{eq:solution-space}, let us denote agent $p_{m}^j$'s known solution space in period $t$ by $\mathbf{S}_{mt}^j$. Then, agents compute their bid, which is the maximum estimated utility they can attain given the known solution space in the current period, according to
     \begin{equation}
         b_{mt}^j =\max_{d^\prime \in \mathbf{S}_{{m}t}^j} U\left(d^\prime, \mathbf{D}_{m\{t-1\}} \right)~.
     \end{equation}
    In line with Eq. \ref{eq:residual-decisions}, $\mathbf{D}_{m\{t-1\}}$ indicates the solutions implemented outside of agent $p_{m}^j$'s subtask in the last period, which agents can infer from $\mathbf{d}_{t-1}$.\footnote{Since agents are myopic, they do not take future pay-offs into account and place their bids only on their immediate utility.} 
     \item Since every agent has the incentive to participate in the auction, $P_m$ bids are submitted by the agents capable of solving the associated subtask for each slot in the group.

     \item The agent who submitted the highest bid for task $\mathbf{d}_m$ wins the auction, joins the group at slot $m$, and gets charged the second-highest bid.\footnote{Auctions in which the top bidder pays the second-highest price are optimal in revealing the bidders' true preferences when the agents' information about other agents' bids is restricted or non-existent \citep{Vickrey1961}. The prices that agents get charged for joining the group are transferred to the organization.} Consequently, the group in that particular period is composed of $M$ agents (one per subtask) who know the solutions that lead to the best (estimated) performance. 
 \end{enumerate}

Figure \ref{fig:group-formation} summarizes the group-formation process. By recurrently holding such auctions, the group adapts its composition so that it best fits the task. Throughout $T$ periods, auctions occur every $\tau$ time steps. Lower (higher) values of $\tau$ indicate a larger (smaller) interval between two auctions and can be interpreted in terms of a long-term (short-term) group composition. An auction always occurs in the first time step, irrespective of the value of $\tau$. We consider three different scenarios:
\begin{itemize}
    \item In the case of \textit{long-term group composition}, the group is formed once in the first period. Only one auction is held. For practical purposes, we refer to this case as $\tau=\emptyset$. 
    \item In the case of \textit{medium-term group composition}, we set auctions taking place every $\tau=10$ time steps. 
    \item To model \textit{a short-term group composition}, we set $\tau=1$, so that auctions occur at every time step. 
\end{itemize}

From an organization's perspective, $\tau$ is a design parameter since by changing the value of $\tau$, organizations can control the lifetime of groups. For example, if $\tau$ is very low (high), an organization gives a group the opportunity of autonomously changing its composition in short (long) intervals. We refer to the actions taken by the organization to reduce the time between auctions (i.e., to decrease $\tau$) as \textit{promoting group adaptation.}

\subsection{Scheduling and parameters for simulation experiments}\label{sec:variables}

\begin{figure}[htp]
    \centering
    \includegraphics[scale=0.4]{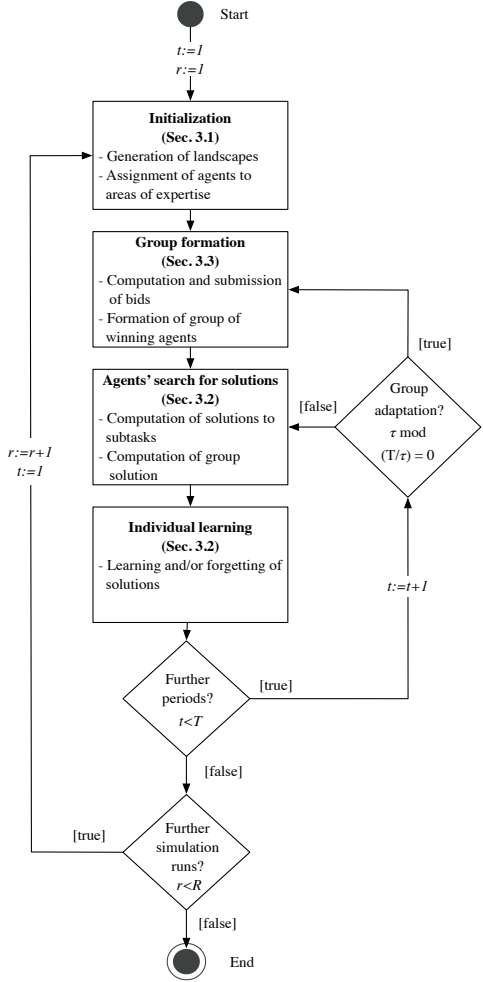}
    \caption{Sequence of events during simulation runs.}
    \label{fig:process}
\end{figure}

We summarize the sequence of events during a simulation run in Fig. \ref{fig:process}. In the previous sections, we have introduced the following independent variables considered in the model: 
\begin{enumerate*}[label=\textit{(\roman*)}]
\item The probability of individual learning $\mathbb{P}$, 
\item the number of auctions $\tau$ during the observation period, and
\item the task complexity $K$.
\end{enumerate*} 

Since organizations can control the learning probability $\mathbb{P}$ and the number of auctions $\tau$ in a real-world setting, we regard those two variables as design parameters. The task complexity $K$ and the structure of interdependencies, on the contrary, are usually given and cannot be controlled by an organization.
All relevant parameters included in the model are summarized in Tab. \ref{tab:variables}. The parameter settings result in a total number of $3 \cdot 2 \cdot 3 = 18$ different scenarios. We set the observation period to $T=200$ and fix the number of subtasks included in the task to $M=3$. We perform $R=1,500$ simulations for every scenario and observe the performance at the group level as the dependent variable.\footnote{The number of simulations was fixed after analyzing the results' variance, as \cite{Lorscheid2012} suggested} The computation of the group's solution and the corresponding performance is formalized in Eqs. \ref{eq:groupsolution} and \ref{eq:perf}, respectively. 
 
\begin{table}[htp]
\scriptsize
\caption{Parameter setting}
\label{tab:variables}
\begin{tabular}{llll}
\multicolumn{1}{l}{Type}            & \multicolumn{1}{l}{Variables} & Notation        & Values \\ 
\noalign{\smallskip} 
\hline 
\noalign{\smallskip} 
\multirow{3}{*}{Independent variables}
    & Interval between auctions   &  $\tau$               & \{$\emptyset$, 1, 10\}\\
    & Task complexity            & $K$                   & \{3, 5\}\\
    & Learning probability  & $\mathbb{P}$          & \{0, 0.25, 0.5\} \\ 
\noalign{\smallskip} \hline \noalign{\smallskip}  
Dependent variable
    & Group performance     & $\bar{C}_t$           & $\left[0,1\right]$\\ 
\noalign{\smallskip} \hline \noalign{\smallskip} 
\multirow{7}{*}{Other parameters}   
    & Time steps            & $t$                   & $\{1,...,200\}$\\
    & Observation period    & $T$                   & 200 \\ 
    & Number of decisions   & $N$                   & 12 \\ 
    & Population of agents  & $P$                   & 30 \\ 
    & Number of subtasks    & $M$                   & 3 \\ 
    & Incentive parameters& $\alpha, \beta$         & 0.5 \\ 
    & Number of simulations & $R$                   & 1,500 \\ 
\noalign{\smallskip} \hline
\end{tabular}
\end{table}

\subsection{Performance measures}\label{sec:observation}

\paragraph{Normalized performance}
To assure that the results are comparable across all simulation runs, we normalize the performance of the group solution by the maximum achievable performance in every landscape. We compute the average normalized performance in period $t$ according to 

\begin{equation}
\label{eq:final-perf}
    \Bar{C}_t = \frac{1}{R} \sum_{r=1}^{R} \frac{C\left( \mathbf{d}_{tr}\right)}{C^{\ast}_r}~,
\end{equation}

\noindent where $C^{\ast}_r$ indicates the maximum performance in simulation run $r$, $\mathbf{d}_{tr}$ stands for the group solution implemented in simulation run $r$ and period $t$. The performance $C\left( \mathbf{d}_{tr}\right)$ is computed according to Eq. \ref{eq:perf}. 

\paragraph{Mean performance}
To give a condensed performance measure, we also report the mean performance over the entire observation period, 
\begin{equation}
\label{eq:mean-perf}
    \Bar{\Bar{C}} = \frac{1}{T} \sum_{t=1}^{T} \bar{C}_t~.
\end{equation}

\noindent Please note that the mean performance includes information about the attainable performance and the speed at which this performance level is reached in a specific scenario. Thus, the mean performance is also a proxy for the convergence speed. 

\paragraph{Interaction effect}

In addition to the two measures of performance introduced above, we aim to analyze the interaction between individual learning and group adaptation when they are promoted simultaneously. To do so, we compute the interaction coefficient based on the final and the mean performance. The analysis starts at a baseline scenario in which neither learning nor group adaptation is promoted. We refer to the parameter set for this case by $\delta_0 = \{\mathbb{P}=0, \tau=\emptyset\}$. To analyze the isolated effects of promoting the design parameters, we start from the baseline scenario and promote either learning or group adaptation, so that the parameters are $\delta_{\mathbb{P}} = \{\mathbb{P}>0, \tau=\emptyset\}$ or $\delta_{\tau} = \{\mathbb{P}=0, \tau>0\}$, respectively. Finally, we analyze the case when learning and group adaptation are promoted simultaneously with the corresponding parameters $\delta_{\mathbb{P}\tau} = \{\mathbb{P}>0, \tau>0\}$.\footnote{For $\mathbb{P}>0$ and $\tau>0$, we consider the values listed in Tab. \ref{tab:variables}.}
We denote the normalized performance in the final period $t=200$ given a specific parameter setting by $\Bar{C}^{\delta}$, where $\delta \in \{\delta_0, \delta_{\mathbb{P}},\delta_{\tau},\delta_{\mathbb{P}\tau}\}$ (see also Eqs. \ref{eq:final-perf} and  \ref{eq:mean-perf}). We suppress the subscript $t=200$ for readability.  

Next, we compute the differences in the achieved final performances. We denote the difference between the final performance achieved in the baseline scenario and the final performance if either individual learning or group adaptation is promoted by $\Bar{\Delta}_0^{{\mathbb{P}}} =  \Bar{C}^{\delta_{\mathbb{P}}} - \Bar{C}^{\delta_0}$ and $\Bar{\Delta}_0^{{\tau}} = \Bar{C}^{\delta_{\tau}} - \Bar{C}^{\delta_0} $, respectively. The increase in the final performance when individual learning and group adaptation are promoted simultaneously is denoted by $\Bar{\Delta}_0^{{\mathbb{P}\tau}} =  \Bar{C}^{\delta_{\mathbb{P}\tau}} - \Bar{C}^{\delta_0}$. Finally, we compute the interaction coefficient by 
\begin{equation}
\label{eq:interaction}
    I\!E=\myfrac[3pt]{\Bar{\Delta}_0^{{\mathbb{P}\tau}}}{\Bar{\Delta}_0^{{\mathbb{P}}}+\Bar{\Delta}_0^{{\tau}}}~.
\end{equation}
Consequently, an interaction coefficient $I\!E>1$ and $I\!E<1$ indicates over- and underproportional interactions, respectively \citep{Leitner2014B}. The effect of simultaneously promoting individual learning and group adaptation on the mean performance is computed correspondingly.

\paragraph{Offsetting effect}
In addition to the interaction between individual learning and group adaptation, we are interested in the potential offsetting effects when the design parameters are changed sequentially. To do so, we analyze the relative difference in the performance of promoting individual learning (group adaptation) after group adaptation (individual learning) has been promoted. 
When individual learning is promoted after group adaptation, we compute the difference in the final performance by
$\Bar{\Delta}_{\tau}^{{\mathbb{P}\tau}}=  \Bar{C}^{\delta_{\mathbb{P}\tau}} - \Bar{C}^{\delta_{\tau}}$. Consequently, the relative difference in the final performance follows 

\begin{equation}\label{eq:complearning}
    O\!E_{\mathbb{P}}=\myfrac[3pt]{\Bar{\Delta}_{\tau}^{{\mathbb{P}\tau}}}{\Bar{C}^{\delta_{\tau}}}~.
\end{equation}
For the case of promoting individual learning first and group adaptation second, we compute the absolute differences in the final performance by 
$\Bar{\Delta}_{\mathbb{P}}^{{\mathbb{P}\tau}}=  \Bar{C}^{\delta_{\mathbb{P}\tau}} - \Bar{C}^{\delta_{\mathbb{P}}}$
and the relative difference in the final performance by 

\begin{equation}\label{eq:compadaptation}
    O\!E_{\tau}=\myfrac[3pt]{\Bar{\Delta}_{\mathbb{P}}^{{\mathbb{P}\tau}}}{\Bar{C}^{\delta_{\mathbb{P}}}}
\end{equation}

\noindent The relative differences in the mean performances are computed correspondingly.

\backmatter

\section{Results}\label{sec:results}

We analyze the effects of an adaptation at two levels - via individual learning and changing group composition - on task performance and report the final and mean performances for decomposed and interdependent tasks in Tab. \ref{tab:performance}.\footnote{Please note that the mean performance is a condensed performance measure that also includes information about the convergence speed.} We organize the results in three subsections: In Sec. \ref{subsec:interaction}, we analyze the interaction effects between individual learning and group adaptation and study how a \textit{simultaneous} variation in both design parameters affects task performance. In Secs. \ref{sec:positive_effects_learning} and \ref{subsec:group-policy}, we focus on a \textit{sequential} promotion of individual learning and group adaptation. In particular, Sec. \ref{sec:positive_effects_learning} analyzes the effects of promoting individual learning after group adaptation has been promoted. Hence, a group's lifetime is fixed before there is any promotion of individual learning. We plot the effects of a subsequent variation in individual learning on task performance in Figs. \ref{fig:result2} and \ref{fig:result3}. Table \ref{tab:learning} includes information about whether or not subsequently promoting learning has significant effects on the mean and final performances. In Sec. \ref{subsec:group-policy}, we explore the effects of promoting group adaptation after learning has been promoted. Figures \ref{fig:result4} and \ref{fig:result5} and Tab. \ref{tab:adaptation} show whether and how a variation in the lifetime of a group after promoting learning affects task performance. 

Our analysis also considers whether different configurations of the incentive systems (i.e., $\alpha$ and $\beta$) and different structures of interdependencies (i.e., Fig. \ref{fig:matrix}) affect the results. In both cases, there are slight differences in the overall performances achieved, but the findings presented in Secs. \ref{subsec:interaction} to \ref{subsec:group-policy} hold true for different configurations of the incentive system. Also, when tasks are interdependent, we observe the same patterns in the performances for different structures of interdependencies. The details are provided in the Appendix.

\begin{table}[!htbp]
\scriptsize
\caption{Mean and final performances}
\label{tab:performance}
\renewcommand{\arraystretch}{1.5}
\resizebox{\textwidth}{!}{%
\begin{tabular}{llcccccc}
                             &             & \multicolumn{3}{c}{Decomposed tasks} & \multicolumn{3}{c}{Interdependent tasks} \\ \cmidrule(l{2pt}r{2pt}){3-5}  \cmidrule(l{2pt}r{2pt}){6-8} 
                             &             & \multicolumn{3}{c}{Learning}         & \multicolumn{3}{c}{Learning}             \\ 
Group composition              & Performance & Zero       & Moderate    & High      & Zero        & Moderate      & High       \\ \hline
\multirow{2}{*}{Long-term}   & Mean        & 0.8687     & 0.9750      & 0.9882    & 0.7524      & 0.9076        & 0.9190     \\
                             & Final       & 0.8697     & 0.9994      & 1.0000    & 0.7529      & 0.9363        & 0.9341     \\ \hline
\multirow{2}{*}{Medium-term} & Mean        & 0.8723     & 0.9915      & 0.9934    & 0.7910      & 0.9227        & 0.9201     \\
                             & Final       & 0.8734     & 1.0000      & 1.0000    & 0.7942      & 0.9411        & 0.9344     \\ \hline
\multirow{2}{*}{Short-term}  & Mean        & 0.8692     & 0.9947      & 0.9963    & 0.7921      & 0.8864        & 0.8680     \\
                             & Final       & 0.8702     & 1.0000      & 1.0000    & 0.7920      & 0.8933        & 0.8708     \\ \hline
\end{tabular}%
}
\end{table}

\subsection{Interactions between promoting individual learning and group adaptation}
\label{subsec:interaction}

\paragraph{Interaction effects}
First, this section analyzes the interaction effects if both individual learning and group adaptation are simultaneously promoted. To do so, we compute the interaction coefficient introduced in Eq. \ref{eq:interaction}. This coefficient gives information about whether promoting learning and group adaptation reinforce or mitigate each other or interact linearly. A reinforcing (mitigating) interaction effect means that the joint effect of simultaneously promoting learning and group adaptation is larger (smaller) than the sum of the isolated effects. Isolated effects refer to the mean and final performance when either individual learning or group adaptation is promoted (see Tab. \ref{tab:performance}). Reinforcing and mitigating effects indicate an over- and underproportional interaction and are indicated by $I\!E>1$ and $I\!E<1$, respectively.  

\begin{figure}[t]
    \centering
    \includegraphics[width=\linewidth]{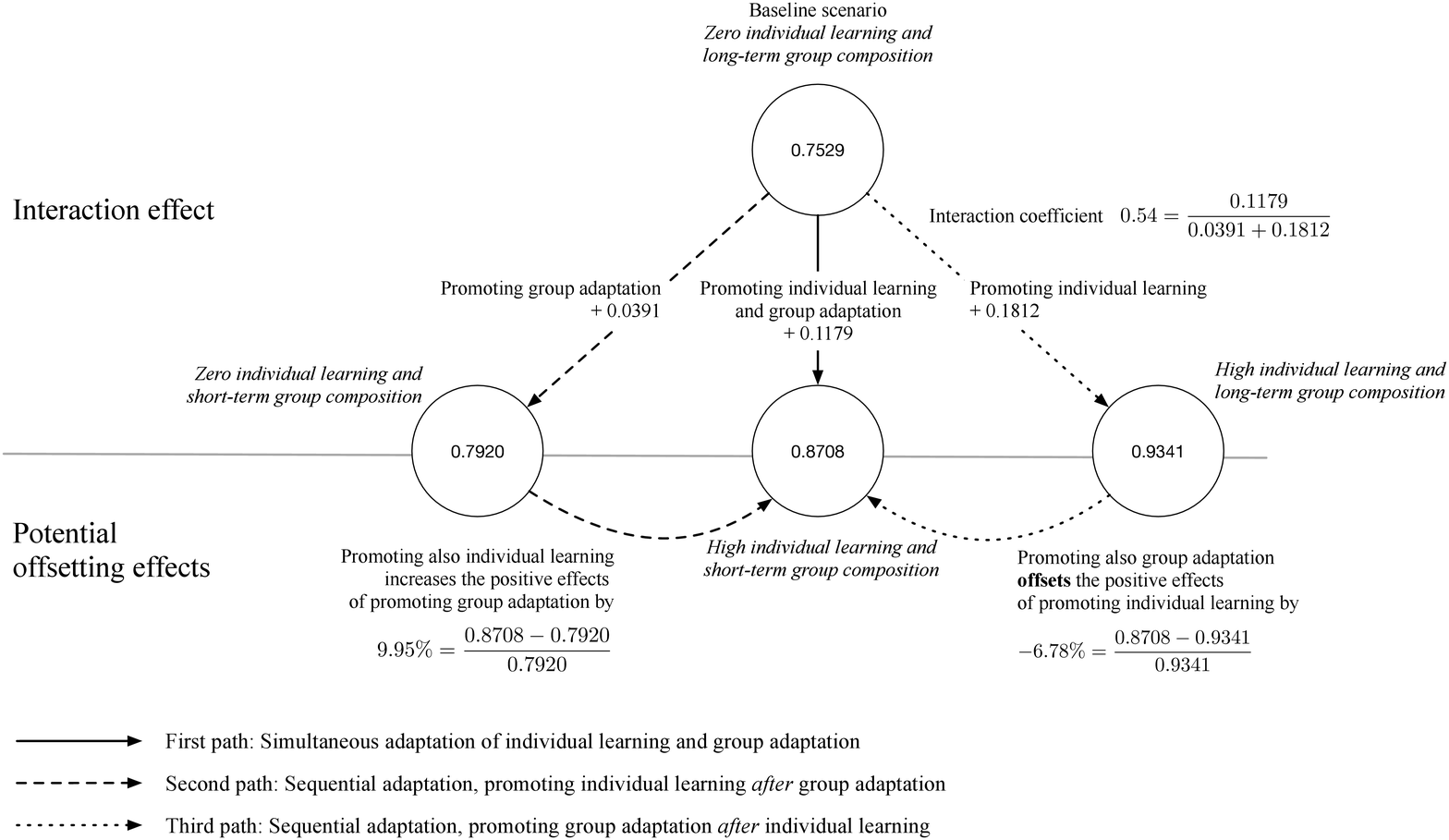}
    \caption{Interaction effect and potential offsetting effects (numbers are for the final performance and interdependent tasks).}
    \label{fig:interactioncoefficient}
\end{figure}

Following Eq. \ref{eq:interaction}, we consider a baseline scenario in which neither individual learning nor group adaptation is promoted (i.e., zero individual learning and long-term group composition). Starting from there, we change both design parameters simultaneously by promoting either learning (moderate or high) or group composition (medium- or short-term). The logic of the interaction coefficient is illustrated in Fig. \ref{fig:interactioncoefficient}, where we refer to one scenario included in Tab. \ref{tab:performance}: In the baseline scenario for an interdependent task with no learning and long-term group composition, a final performance of $0.7529$ can be achieved. Promoting group adaptation towards a short-term composition (individual learning towards a high probability) increases this performance by $0.0391$ ($0.1812$). However, when both design parameters are changed at a time, the performance increases by $0.1179$; see the first path indicated by the solid line in Fig. \ref{fig:interactioncoefficient}. The resulting interaction coefficient is $I\!E=0.54$ and suggests that the joint effect is smaller than the sum of the isolated effects. Thus, the effects of promoting individual learning and group adaptation \textit{simultaneously} interact underproportionally in this case. 

The interaction coefficients for decomposed and interdependent tasks are presented in Tab. \ref{tab:interactions}. When tasks are decomposed, the effects of promoting individual learning and group adaption interact more or less linearly.  We observe a slight overproportional effect on the mean performance only if the probability of individual learning increases to $0.25$ with a simultaneous promotion of group adaption to either a medium-term or short-term composition. By contrast, when tasks are interdependent, the results indicate underproportional interactions in all cases. This means that the joint effect of simultaneously promoting individual learning and group adaptation is smaller than the sum of the isolated effects.  

\begin{table}[!htbp]
\caption{Interaction coefficients}
\label{tab:interactions}
\renewcommand{\arraystretch}{1.5}
\resizebox{\textwidth}{!}{%
\begin{tabular}{llcccc}
                                &             & \multicolumn{2}{c}{Decomposed tasks}                                    & \multicolumn{2}{c}{Interdependent tasks}                                \\ \cmidrule(l{2pt}r{2pt}){3-4}
                                \cmidrule(l{2pt}r{2pt}){5-6}
                                &             & \multicolumn{2}{c}{Learning}                                            & \multicolumn{2}{c}{Learning}                                            \\ 
Group composition                 & Performance & \multicolumn{1}{l}{Zero to moderate} & \multicolumn{1}{l}{Zero to high} & \multicolumn{1}{l}{Zero to moderate} & \multicolumn{1}{l}{Zero to high} \\ \hline
\multirow{2}{*}{Long-term to medium-term} & Mean        & 1.12                                 & 1.01                             & 0.88                                 & 0.82                             \\
                                & Final       & 0.98                                 & 0.97                             & 0.84                                 & 0.82                             \\ \hline
\multirow{2}{*}{Long-term to short-term}  & Mean        & 1.18                                 & 1.06                             & 0.69                                 & 0.56                             \\
                                & Final       & 1.00                                 & 1.00                             & 0.63                                 & 0.54                             \\ \hline
\end{tabular}%
}
\end{table}

\paragraph{Potential offsetting effects}
Since there are underproportional interactions in all scenarios with interdependent tasks, it would be interesting to know whether and, if so, to what extent promoting individual learning offsets the increase in performance produced by promoting group adaptation and vice versa.\footnote{Recall, we observe almost linear interactions for decomposed tasks which is why no offsetting effects are expected in these cases.} For this purpose, we consider two-stage paths that promote individual learning and group adaptation \textit{sequentially in a different order} and compute the relative changes in performance in the second stage (see Eqs. \ref{eq:complearning} and \ref{eq:compadaptation}). The logic behind our analysis is illustrated in Fig. \ref{fig:interactioncoefficient}. Again, we start at the baseline scenario with a performance of $0.7529$. The first path, indicated by a solid line, captures the interaction effect as described above. The second path, indicated by dashed lines, considers promoting group adaptation first, resulting in a performance of $0.7920$, and individual learning second. In the second stage, the performance increases by $9.95\%$ to $0.8708$ (see Tabs. \ref{tab:performance} and \ref{tab:complearn}). Thus, promoting individual learning after promoting group adaptation does \textit{not} offset any positive effects in this scenario. The third path, indicated by dotted lines in Fig. \ref{fig:interactioncoefficient}, considers promoting individual learning first, resulting in a performance of $0.9341$, and group adaptation second. In the second stage, the performance \textit{decreases} by $6.78\%$ to $0.8708$ (see Tabs. \ref{tab:performance} and \ref{tab:compgroup}). Thus, in this path, promoting group adaptation in the second stage offsets the performance increase achieved by promoting individual learning at the first stage. 

\begin{table}[!htbp]
\scriptsize
\caption{Relative performance changes caused by promoting individual learning (interdependent tasks).}
\label{tab:complearn}
\renewcommand{\arraystretch}{1.5}
\begin{tabular}{llcc}
                             
                             &              & \multicolumn{2}{c}{Learning}             \\ 
Group composition            & Performance  & Zero to Moderate           & Zero to High              \\ \hline
\multirow{2}{*}{Long-term}   & Mean         & 20.63\%               & 22.15\%            \\
                             & Final        & 24.37\%               & 24.07\%            \\ \hline
\multirow{2}{*}{Medium-term} & Mean         & 16.65\%               & 16.32\%            \\
                             & Final        & 18.49\%               & 17.64\%            \\ \hline
\multirow{2}{*}{Short-term}  & Mean         & 11.90\%               & 9.57\%            \\
                             & Final        & 12.79\%               & 9.95\%            \\ \hline
\end{tabular}%
\end{table}

Table \ref{tab:complearn} shows the relative changes for promoting individual learning in the second stage when tasks are interdependent. There are no offsetting effects but, conversely, further increases in the final performance. This observation is most pronounced if groups do not adapt at all (i.e., $24.37\%$ and $24.07\%$) and least pronounced if groups change their composition in the short-term (i.e., $12.79\%$ and $9.95\%$). 
However, these results also show that it still pays off to promote individual learning in the second stage but, at best, only up to a moderate level because any further increase in individual learning would reduce the final performance: Promoting learning from zero to moderate results in an increase of $12.79\%$, whereas promoting from zero to high leads to an increase of only $9.95\%$. 

\begin{table}[!htbp]
\scriptsize
\caption{Relative performance changes caused by promoting group adaptation (interdependent tasks).}
\label{tab:compgroup}
\renewcommand{\arraystretch}{1.5}
\begin{tabular}{llcc}
                          &             & \multicolumn{2}{c}{Group Composition} \\ 
Learning                  & Performance & Long-term to medium-term     & Long-term to short-term    \\ \hline
\multirow{2}{*}{Zero}     & Mean        & 5.13\%              & 5.28\%            \\
                          & Final       & 5.49\%              & 5.20\%            \\ \hline
\multirow{2}{*}{Moderate} & Mean        & 1.66\%              & -2.33\%           \\
                          & Final       & 0.51\%              & -4.60\%           \\ \hline
\multirow{2}{*}{High}     & Mean        & 0.12\%              & -5.55\%           \\
                          & Final       & 0.03\%              & -6.78\%           \\ \hline
\end{tabular}%
\end{table}

Table \ref{tab:compgroup} shows the relative changes for promoting group adaptation in the second stage when tasks are interdependent. The performance only slightly increases if there is no individual learning. At higher levels of individual learning, promoting group adaptation offsets  preceding positive effects so that the performance achieved in the first stage even decreases. In particular, in the case of no individual learning, changing a group's composition is somewhat beneficial to the performance in the medium-term instead of not changing it at all (i.e., $5.49\%$). The marginal effects of promoting a short-term instead of a medium-term group composition are negligible since the final performances almost do not differ (i.e., $5.49\%$ and $5.20\%$). This means, promoting group adaptation towards a short-term composition does not pay off even if individuals do not learn at all. Given a moderate or high level of individual learning, there is no effect on performance when changing from long-term to medium-term group composition in the second stage (i.e., $0.51\%$ and $0.03\%$). Promoting group adaptation towards a short-term composition even decreases the final performance (i.e., $-4.60\%$ and $-6.78\%$).

\subsection{Promoting individual learning after promoting group adaptation}
\label{sec:positive_effects_learning}

The results presented in Tab. \ref{tab:performance} suggest that the performance always increases as agents start to learn at the second stage, i.e., when the learning probability increases from $\mathbb{P}=0$ to $\mathbb{P}=0.25$.
This finding follows intuition and can be observed for decomposed and interdependent tasks. Since learning enables agents to find new and perhaps better-performing solutions to their subtasks, performance may steadily increase. However, after a certain number of periods, the performance cannot be further improved and settles at a specific level, which we refer to as the \textit{limit performance} (i.e., the maximum performance that agents can achieve \textit{on average}, given the conditions considered in the scenario). The limit performance increases significantly with the learning probability.
While both the final and mean performances increase if agents only start to learn, further increases in the learning probability lead to more differentiated effects that require taking into account the lifetimes of groups.

\begin{figure}[!tp]
    \centering
    \includegraphics[width=\linewidth]{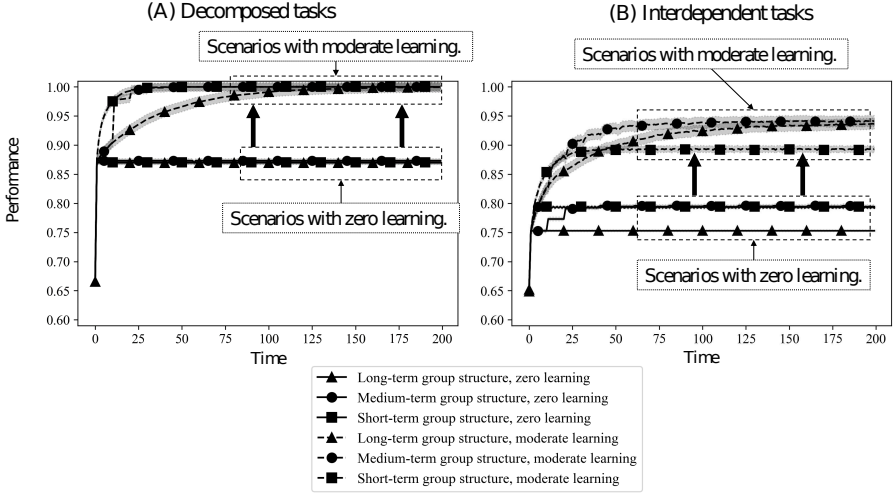}
    \caption{Effects of promoting individual learning in the case of a long-term group composition.}
    \label{fig:result2}
\end{figure}

\paragraph{Long-term group composition} In the case of promoting a long-term group composition at the first stage, the convergence speed rises if the learning probability increases from moderate ($\mathbb{P}=0.25$) to high ($\mathbb{P}=0.5$). However, an increase in the learning probability to values above $0.25$ has no significant effect on the final performance. The results included in Tabs. \ref{tab:complearn} and \ref{tab:learning} confirm this finding, and Fig. \ref{fig:result2} shows the same pattern for decomposed and interdependent tasks. 

\begin{table}[!htbp]
\caption{Significance test for the effects of promoting individual learning at the second stage}
\label{tab:learning}
\renewcommand{\arraystretch}{1.5}
\resizebox{\textwidth}{!}{%
\begin{tabular}{llcccc}
                             &                        & \multicolumn{2}{c}{Decomposed tasks}    & \multicolumn{2}{c}{Interdependent tasks} \\ \cmidrule(l{2pt}r{2pt}){3-4} \cmidrule(l{2pt}r{2pt}){5-6} 
                             && \multicolumn{2}{c}{Learning} & \multicolumn{2}{c}{Learning}\\
Group composition              & Performance            & Zero to moderate     & Moderate to high & Zero to moderate    & Moderate to high   \\ \hline
\multirow{2}{*}{Long-term}   & Mean & **                   & **               & **                  & **                 \\
                             & Final  & **                   & n.s.             & **                  & n.s.               \\ \hline
\multirow{2}{*}{Medium-term} & Mean & **                   & n.s.             & **                  & n.s.               \\
                             & Final & **                   & n.s.             & **                  & n.s.               \\ \hline
\multirow{2}{*}{Short-term}  & Mean  & **                   & n.s.             & **                  & **                 \\
                             & Final & **                   & n.s.             & **                  & **                 \\ \hline
\end{tabular}%
}
**  Indicates significance at the 99\% level\\
n.s. Indicates not significant
\end{table}

\begin{figure}[!tp]
    \centering
    \includegraphics[width=\linewidth]{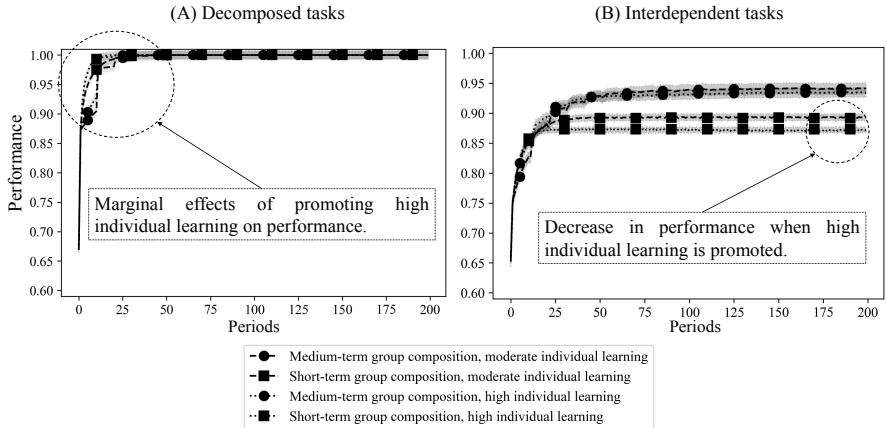}
    \caption{Effects of promoting individual learning in the case of a medium-term and short-term group composition.}
    \label{fig:result3}
\end{figure}

\paragraph{Medium-term and short-term group composition}
For the case of promoting a medium-term group composition at the first stage, there are no moderating effects of task decomposability and no significant effects of increasing the learning probability to $0.5$ on the final and the mean performances (see Tab. \ref{tab:learning}). This corresponds to the results presented in Sec. \ref{subsec:interaction} and in particular in Tab. \ref{tab:complearn}. There are marginal effects of an increase in the learning probability on the speed of performance improvement only in very early periods (see Fig. \ref{fig:result3}(A)).

For the case of promoting a short-term group composition at the first stage, the results included in Fig. \ref{fig:result3} and Tab. \ref{tab:learning} indicate that the effects of promoting learning are moderated by task decomposability. There is a significant decrease in the final and mean performance when tasks are interdependent. This result is in line with the decrease in the relative performance changes presented in Tab. \ref{tab:complearn} (e.g., $12.79\%$ to $9.95\%$ for the final performance in the case of a short-term group composition). Further, there are no significant effects of promoting learning on performances when tasks are decomposed, reflecting the linear interactions presented in Tab. \ref{tab:interactions}. 

\subsection{Promoting group adaptation after promoting individual learning}
\label{subsec:group-policy}

While promoting individual learning at the second stage has significant consequences for the performance in most cases, promoting group adaptation after promoting individual learning has less pronounced effects. 

\begin{figure}[!tp]
    \centering
    \includegraphics[width=\linewidth]{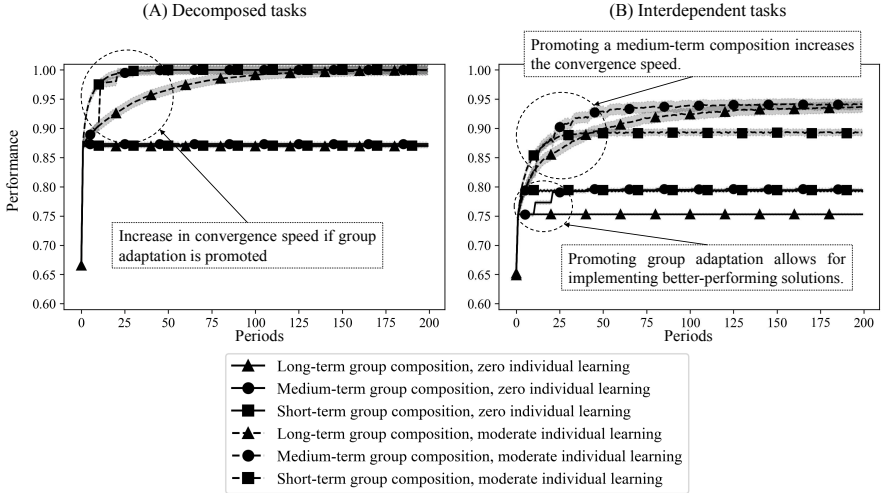}
    \caption{Effects of increasing group adaptation in the case of no and moderate individual learning.}
    \label{fig:result4}
\end{figure}

\paragraph{No individual learning}
In the case of not promoting individual learning at the first stage, the results included in Fig. \ref{fig:result4}(B) and Tabs. \ref{tab:performance} and \ref{tab:adaptation} show that groups of a short-term or medium-term composition achieve significantly higher performances than groups that do not adapt at all. Still, this effect can only be observed when tasks are interdependent. Thus, there is a moderating effect of task decomposability. The observation that promoting group adaptation increases performance follows intuition: When individuals do not learn and a group has the opportunity to change its composition in the short- or medium term, new group members may know better-performing solutions to the task at hand and, consequently, performance may increase. Since groups that do not adapt at all cannot acquire knowledge by attracting new members, the increase in task performance comes to a standstill at a low level early in the observation period. The effects on performance, however, are relatively small. Promoting a short-term instead of a medium-term composition has no significant effect on the final performance, if agents do not learn at the individual level. Yet, the level of the final performance is achieved slightly faster (see Fig.\ref{fig:result4}(B)). This is also reflected in the results presented in Sec. \ref{subsec:interaction} and in particular in Tab. \ref{tab:compgroup}.

\begin{table}[!htbp]
\caption{Significance test for the effects of promoting group adaptation at the second stage}
\label{tab:adaptation}
\renewcommand{\arraystretch}{1.5}
\resizebox{\textwidth}{!}{%
\begin{tabular}{llcccc}
                          &                        & \multicolumn{2}{c}{Decomposed tasks}   & \multicolumn{2}{c}{Interdependent tasks} \\ \cmidrule(l{2pt}r{2pt}){3-4} \cmidrule(l{2pt}r{2pt}){5-6} 
                           && \multicolumn{2}{c}{Group composition} & \multicolumn{2}{c}{Group composition}\\
Learning                 & Performance              & \Centerstack{Long-term to \\ medium-term}       & \Centerstack{Medium-term to \\ short-term} &  \Centerstack{Long-term to \\ medium-term}     & \Centerstack{Medium-term to \\ short-term}        \\ \hline
\multirow{2}{*}{Zero}     & Mean  & n.s.                 & n.s.            & **                & n.s.                \\
                          & Final & n.s.                 & n.s.            & **                 & n.s.                \\ \hline
\multirow{2}{*}{Moderate} & Mean  & **                   & n.s.            & **                 & **                  \\
                          & Final & n.s.                 & n.s.            & n.s.               & **                  \\ \hline
\multirow{2}{*}{High}     & Mean & n.s.                 & n.s.            & n.s.               & **                  \\
                          & Final & n.s.                 & n.s.            & n.s.               & **                  \\ \hline
\end{tabular}%
}
** Indicates significance at the 99\% level\\
n.s. Indicates not significant
\end{table}

\paragraph{Moderate individual learning}
When moderate individual learning is promoted at the first stage, groups of a medium-term composition improve their performance significantly faster than groups that do not adapt at all. However, there are no significant differences in the achieved final performances (see Tab. \ref{tab:adaptation}). While the same result can be observed for both interdependent and decomposed tasks, task decomposability appears to have a moderating effect on the convergence speed and the final performance when group adaption is further promoted towards a short-term composition. For interdependent tasks, the results presented in Fig. \ref{fig:result4}(B) indicate that groups that are of a short-term composition achieve a significantly lower performance than groups that adapt less frequently, i.e., are of a medium-term composition (see Tab. \ref{tab:performance}). That means, too much group adaptation might unfold adverse effects on performance (see also the negative values reported in Tab. \ref{tab:compgroup} in Sec. \ref{subsec:interaction}).

\begin{figure}[!tp]
    \centering
    \includegraphics[width=\linewidth]{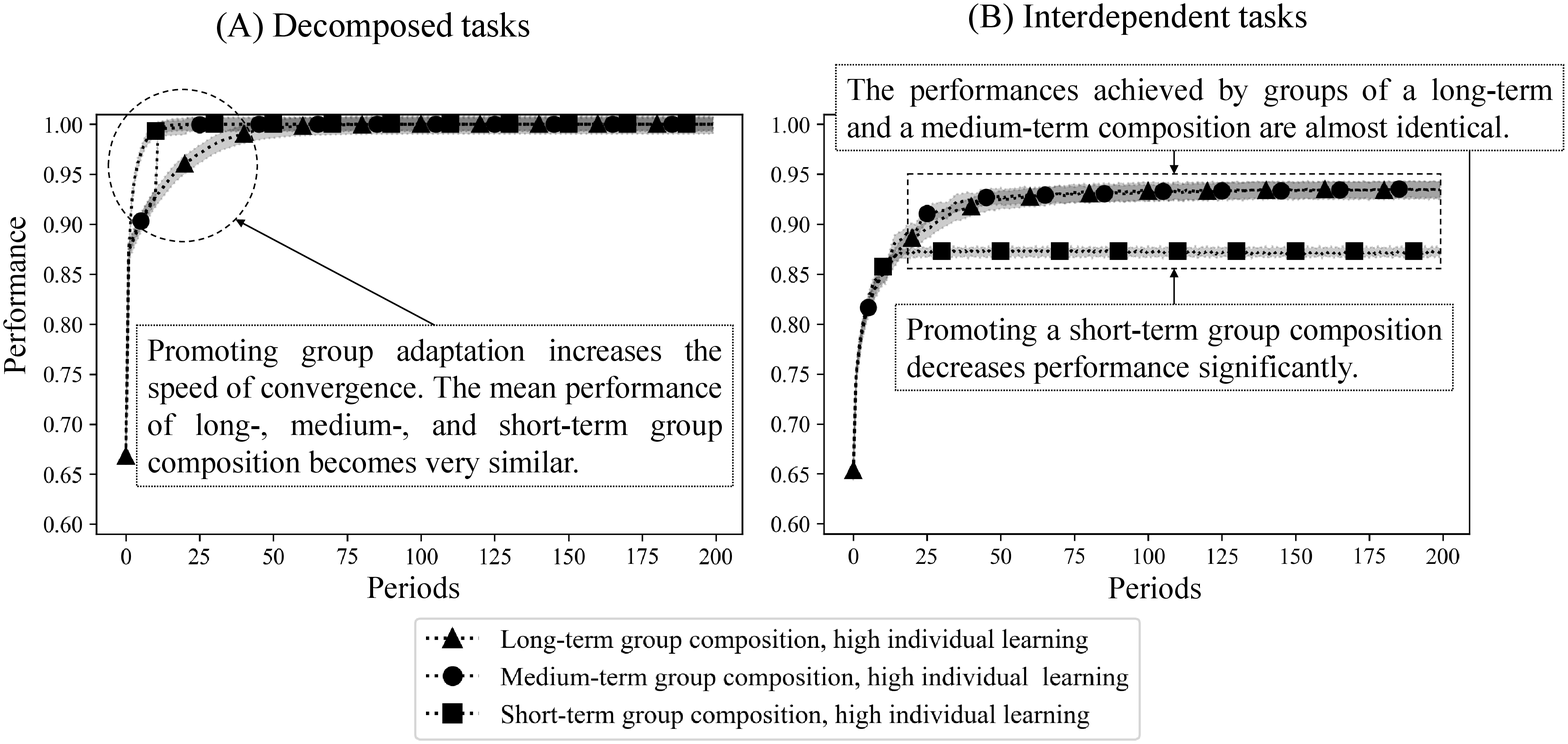}
    \caption{Effects of promoting group adaptation in the case of high individual learning.}
    \label{fig:result5}
\end{figure}

\paragraph{High individual learning}
If high individual learning is promoted at the first stage, the effects of promoting group adaptation are similar to those identified for moderate individual learning. However, if the probability of individual learning is high as compared to lower learning probabilities the performance increases faster when groups are of a long-term composition (see Fig. \ref{fig:result5}). For decomposed tasks, the mean performances achieved by groups of a short-, medium- and long-term composition become even more similar than in the case of moderate learning (see Tab. \ref{tab:performance}). Thus, the benefits of promoting a relatively short-term group composition decrease. If tasks are interdependent, groups of a short-term composition perform significantly worse than groups of a longer lifetime (see Tabs. \ref{tab:performance} and \ref{tab:adaptation}). In contrast, the performances achieved by groups of a long-term and a medium-term composition almost do not differ (see Fig. \ref{fig:result5}(B)). Like for moderate individual learning, task decomposability has a moderating effect when a group's lifetime is short. Promoting group adaptation is not necessarily advantageous if individual learning is already high but might even unfold adverse effects, particularly when tasks are interdependent.

\section{Discussion}\label{sec:discussion}

Our research aims to gain insights into the interaction effects of variations in two design parameters, namely
\begin{enumerate*}[label=\textit{(\roman*)}]
\item individual learning and
\item group adaptation.
\end{enumerate*}
In particular, we aim to understand how the simultaneous and sequential promotion of individual learning and group adaptation affects task performance and how task decomposability moderates any effects. To do so, we have extended the \textit{NK}-framework by a learning mechanism and a group formation mechanism to account for adaptation at the level of the individual agent and the group of agents, respectively. 

\subsection{Results related to interaction effects}\label{subsec:discussion_interaction}

The results presented in this paper can be related to the exploration-exploitation dilemma, which is concerned with the trade-off between obtaining new knowledge and using the available knowledge to improve performance \citep{Berger-Tal2014}. Previous research points out that the key to improving performance is a proper balance between exploration and exploitation \citep{Berger-Tal2014,Levinthal1997,Yen2002}. Further factors, such as the managerial initiative \citep{Podolny2018}, feedback \citep{Giannoccaro2019,Hakonsson2016}, information about the environment \citep{Leitner2020}, task complexity \citep{Uotila2017}, and organizational policies \citep{Staber2002} need to be taken into account.

The previous literature on the topic often employs a unidimensional perspective, in which exploration and exploitation occur either at the individual \citep{Giannoccaro2019,Hakonsson2016,Leitner2020,Podolny2018} or group level \citep{Staber2002,Uotila2017}. Please note that, from the perspective of the organization, promoting individual learning and promoting group adaptation are key design parameters to control whether groups lean more towards exploration or exploitation. Naturally, promoting high learning motivates agents to obtain new knowledge and, thereby, exploration is fostered at the individual level. The group formation mechanism implemented in our model makes sure that the agents who have the best knowledge to solve the task join forces in a group. Thus, promoting group adaptation can be interpreted as a mechanism to foster exploration from a group's perspective. The shorter (longer) the group's lifetime, the more (less) often this mechanism is carried out, and, consequently, the more exploration (exploitation) is promoted. We contribute to this stream of literature by analyzing the interactions between individual learning and group adaptation and quantifying them in terms of the interaction coefficient. 

In Sec. \ref{subsec:interaction}, we have shown that there are not just linear but also non-linear interactions if individual learning and group adaptation are promoted simultaneously. While these interactions are close to linear in all cases for decomposed tasks, we find (highly) underproportional interactions in interdependent tasks. Previous research has addressed the interaction between the effects of promoting learning and group adaptation, too. \citet{Savelsbergh2015}, for example, found a positive relationship between a long-term group composition and team learning. We show that a similar relation also holds true for individual learning. Moreover, \citet{Bartsch2013}, \citet{Edmondson2003}, and \citet{Sergeeva2018} claim that the promotion of a more frequent group adaptation might offset the effects produced by an increase in individual learning. This finding is in line with our observations, and we contribute to this stream of research by showing that this offsetting indeed occurs, but just if tasks are interdependent. 

Moreover, previous research suggests that interactions across subtasks require a broader exploration to develop well-performing solutions to tasks, mainly when landscapes are characterized by a relatively high number of peaks \citep{Rivkin2007}. This is in line with our findings since both promoting individual learning and group adaptation particularly pay off in interdependent tasks. However, we also show that overpromoting exploration might fire back via performance. Thus, we regard it highly important, particularly for corporate practice, \textit{not} to overpromote learning and group adaptation. \citet{Billinger2014}, for example, find that human decision makers are indeed prone to overexploration. We show that cheering this tendency might unfold unwanted behavioral dynamics and might result in a decreasing performance.

\subsection{Results related to promoting individual learning}\label{subsec:discussion_learning}
In Sec. \ref{sec:positive_effects_learning}, we find that starting to learn at the second stage, i.e., after group adaptation has been promoted before, increases performance in all cases. In contrast, the effects of promoting high learning depend on a group's lifetime. Also, for higher levels of learning, the effects are moderated by task decomposability. These results relate to previous research on the relationship between means to promote individual learning and organizational performance measures, such as productivity and financial and innovative performance. These means include mentoring \citep{Lankau2007}, allowing for \enquote*{learning by doing} \citep{Arrow1971} and exploratory learning \citep{Beugelsdijk2008}, determining the flow of information that surrounds learning \citep{Cohen1991}, and, more generally, creating work environments that stimulate creativity and learning \citep{Annosi2020,Oldham2003,Sung2014}. 
In addition, \citet{Stinchcombe1990}, \citet{Cohen1991}, \citet{Salas1999}, and \citet{Tharenou2007} highlight the importance of employee training to promote individual learning since the agents' skills are the foundation of organizational capabilities and, hence, employee training contributes substantially to organizational competitiveness. In this regard, it is important to note that providing training to promote individual learning is widely employed in corporate practice. This can -- not least -- be seen in the enormous amounts of money spent on it \citep{Haccoun1998}. 

Previous research on promoting learning through training mainly focuses on the individual-level outcomes in terms of what was learned by the agent, while the consequences of training for the organizational level is seldom in focus and, consequently, still needs to be explored \citep{Glaveli2011,Kozlowski2000}. The first large-scale studies concerned with the link between training and performance are those carried out by \citet{Holzer1993} and \citet{Bartel1994}. They found evidence for a direct positive relationship between learning and productivity, which is in line with our finding that performance increases if agents start to learn. 

\citet{Becker1975} distinguishes between general and specific training. Specific training exclusively increases the performance of a particular organization, makes new employees familiar with the organization, and helps gain new knowledge in monopolistic environments, where no other organizations exist for which the knowledge would be useful. By contrast, in the case of general training, the knowledge gained might be useful also for competitors \citep[see also][]{Barrett2001}. \citet{Becker1975} claims that most training is neither purely general nor completely specific. We argue that the learning included in our model is of a general nature since it is concerned with how to carry out a task that similar organizations could face. In our model, specific knowledge could be the agents' knowledge about the functioning of the auction mechanism for group adaptation. We assume specific knowledge to be given and, hence, do not focus on it. The emphasis on general training allows us to connect our results to the literature on gift-exchange: As soon as agents realize that the knowledge gained from promoting learning might be useful in other employments as well, they might regard it as a \enquote*{gift} \citep{Akerlof1982,Barrett2001}. Following gift-exchange models \citep{Duffy2014}, employees would eventually repay the gift in one or the other form. In an organizational setting, this repay could take the form of putting more effort into solving the task at hand, which might result in higher performance \citep{Cropanzano1997,Falk2006}. Thus, for the positive relationship between learning and performance observed in our model, the theory of gift-exchange explains a similar pattern. However, our agents act utility maximizing under bounded rationality, which apparently leads to the same patterns at the macro-level. Moreover, it is well known that increasing the \enquote*{gift} does not necessarily lead to agents making more effort since the marginal effect decreases. For the context of monetary incentives, this observation has been explained by a crowding-out effect of rewards or individual earnings targets that pose an upper limit on effort \citep{Camerer1997,Frey1997}. We observe that performance does not increase but perhaps even decreases if learning is promoted too intensely. In these cases, the agents' behavior can, thus, be described by a decreasing marginal \enquote*{gift}-effort relation. It is, however, not a crowding-out effect or a compensation target that drives our observation, but the maximum attainable performance. If agents learn with a high probability, they can achieve this performance faster. However, any further promotion of learning does not pay off because there is no more  room for further improvement. 

\citet{Glaveli2011} argue that, in particular, the factors that mediate the outcomes of promoting learning by training have not yet been substantively explored. We contribute in this respect by showing that promoting learning for untrained individuals yields positive effects in all cases. This finding contrasts that by \citet{Barrett2001}, who found that the positive effects of general training are robust against corporate restructuring. In the context of our model, corporate restructuring can be translated to the design parameter of group adaptation. We contribute to this line of research by showing that task interdependence and promoting group adaptation impair the positive effects of training on performance as soon as agents learn with a higher probability; in the worst cases, learning can even yield negative effects for performance. 

\subsection{Results related to promoting group adaptation}\label{subsec:discussion_group adaptation}

In Sec. \ref{subsec:group-policy}, we have shown that promoting group adaptation at the second stage, i.e., after individual learning has been promoted, particularly pays off when tasks are interdependent. In this case, both the final and mean performances increase. By contrast, for decomposed tasks, there are, at best, increases in the convergence speed.  

Previous research asks to take into account temporal aspects of a group's composition. \citet{Mathieu2014}, for example, argue that including the aspect of time into group composition research allows, amongst others, to model how teams move through a lifecycle from birth to death, temporal norms, and the future orientation of the organizational culture \citep{Mohammed2008}. This is also particularly relevant for organizational research, as \citet{Tannenbaum2012} and \citet{Bell2017} claim, because organizations more frequently keep relying on team-based structures. This means that groups are formed for a predetermined time to solve a specific set of tasks \citep{Lundin1995}. In our research, we account for the lifetime of groups by the decision about whether or not to promote a long-term, medium-term, or short-term group composition. While previous research argues that organizations might strategically use a limited lifetime of groups because this allows them to redeploy their human capital \citep{Bell2017}, our approach is different. We follow an (evolutionary) bottom-up approach of group formation \citep{Tsoukas1993} that is driven by a second-price auction \citep[see, e.g.,][]{Leitner2021} to assure that the best-prepared agents join forces in a group. Previous research, in contrast, sometimes appears to stick to the concept of more classical top-down approaches to group composition \citep{Romme2003}. 
Furthermore, it has already been argued that promoting group adaptation can have different consequences. It either increases the performance because it stimulates creativity within a group \citep{Choi2005} or it decreases performance because newly formed groups require some time to develop efficient modes of collaboration \citep{Lewis2007}. We contribute to this line of research by shedding light on the interactions between the effects of promoting individual learning and group adaptation and by exploring the moderating role of task decomposability. In addition, we show that only the convergence speed but not the final performance might increase in the case of decomposed tasks, which supports the argument brought forward in \cite{Choi2005}. On the contrary, we indeed observe decreases in the final performance if groups change their composition too frequently and if tasks are  interdependent. This insight supports the claim by \citet{Lewis2007}. However, we add that it is not only the initial phase of coalescing that might decrease the performance, but also over-promoted exploration (and the interaction with promoting individual learning) might lead to significant decreases in performance. 

\section{Conclusion}\label{sec:conclusion}

In this paper, we analyze and discuss how learning and adaptation at multiple levels in an organization affect task performance. We contribute to previous research by extending the traditionally unidimensional perspective on either the individual or the group level and exploring the effects when individual learning and group adaptation, in the sense of changing a group's composition, occur simultaneously and sequentially, respectively. Our results indicate that, in general, organizations are well-advised to promote learning and adaptation to increase task performance. However, we also show that individual learning and group adaptation should not be pushed too far because there are interaction effects between the two levels. Whether or not interactions are close to linear is moderated by task decomposability. 
In particular, if the group members are very much engaged in learning at the individual level, changing the group composition may backfire and even decrease performance, at least when tasks are interdependent. If individuals only learn to a rather minor extent, changing the composition of a group from time to time is beneficial to task performance. Still, very short intervals between group adaptations do not pay off. 
However, if organizations allow groups to change their composition in the short term, learning at the individual level is in general beneficial. However, the marginal effects of pushing learning beyond a moderate level are negligible. Our results shed new light on the consequences of simultaneously or sequentially promoting individual learning and group adaptation for performance. By revealing the effects of micro-level activities on macro-level performance we contribute to closing the still predominant gap between micro- and macro-level research in managerial science.

Our research, of course, is not without its limitations. First, we assume that the agents are heterogeneous concerning their capabilities and limited in their rationality. We do not consider the effects of any other individual characteristics -- such as social background, age, gender, education, or national culture -- on performance. Further research may want to elaborate on this. Second, we omit communication and coordination between agents, and we exclusively focus on one group within an organization. Further research could extend our approach by adding communication channels between agents and analyzing the co-evolution of multiple (potentially interdependent) groups. Third, we do not take into account the potential costs of individual learning and group adaptation. Finally, we fix the probability of individuals learning and the intervals at which groups may change their composition exogenously. Future research could consider some self-control and self-generated initiative by individuals and groups and investigate how endogenous decisions on the learning probability and the lifetime of groups affect the results.

\section*{Availability of data}
Simulation data is available  \href{here.}{https://gitlab.aau.at/dablancofern/nk-model-multilevel-adaptation/-/blob/data/Results_-_simulations.xlsx}

\section*{Code availability}
The code is available \href{here.}{https://gitlab.aau.at/dablancofern/nk-model-multilevel-adaptation/-/tree/master}

\newpage
\begin{appendices}
\renewcommand\thefigure{\thesection.\arabic{figure}}    
\setcounter{figure}{0}  

\renewcommand\thetable{\thesection.\arabic{table}}    
\setcounter{table}{0}  

\section{Alternative structure of interdependencies}\label{secA1}

\begin{figure}[!htbp]
 \centering
 \includegraphics[width=0.85\textwidth]{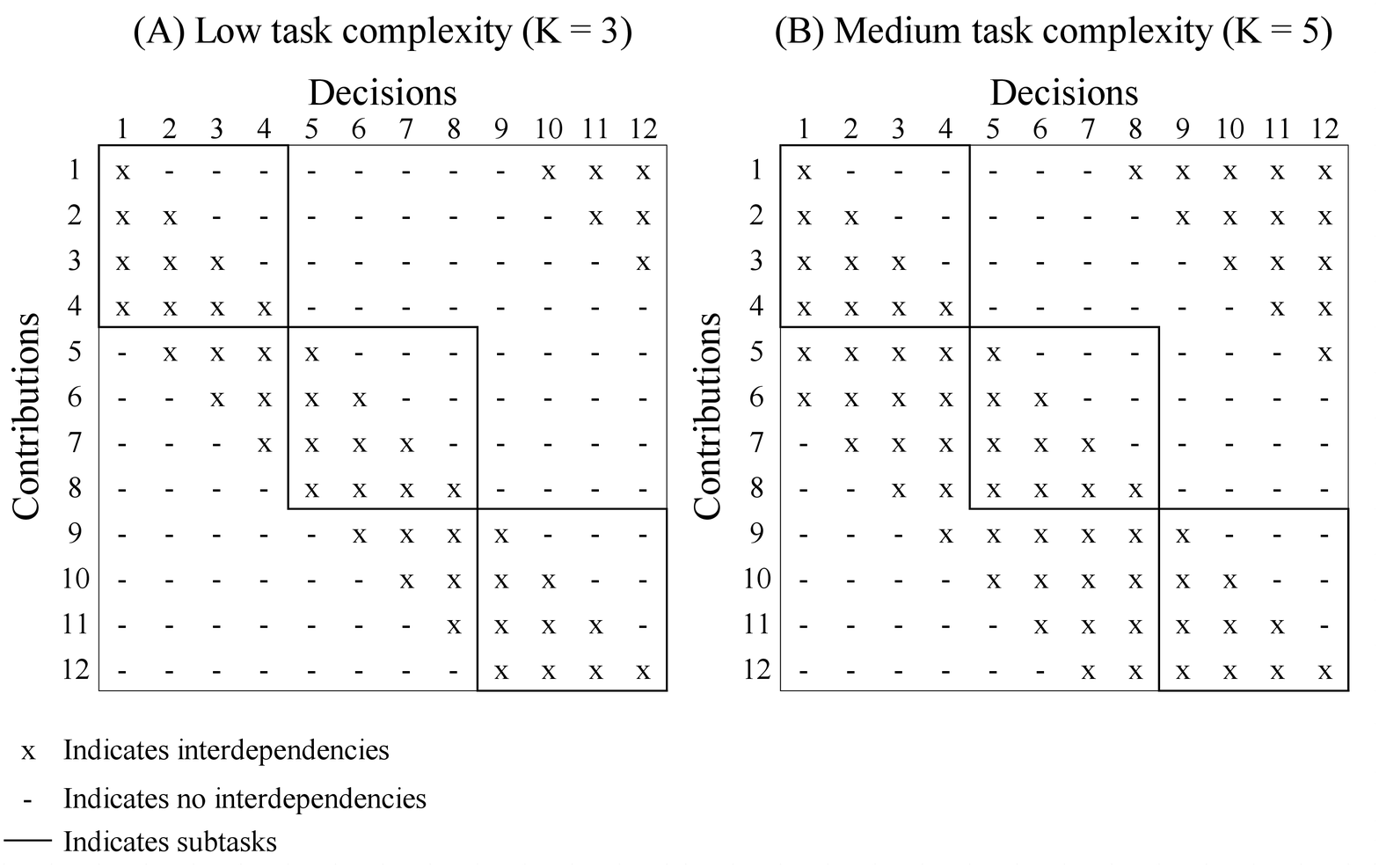}
 \caption{Interdependence matrices for a roll structure.}
 
 \label{fig:altmatrix}
\end{figure}

Figure \ref{fig:altmatrix} illustrates the interdependencies and subtasks for an alternative structure of interdependencies, e.g., a \textit{roll} structure. The interdependencies are partly located outside the subtasks, irrespective of the level of task complexity $K$. 

Table \ref{tab:performanceroll} shows the mean and final performances for the roll structure. As compared to Tab. \ref{tab:performance}, the mean performances are lower for the case of a roll structure. 

\begin{table}[!htbp]
\caption{Mean and final performances for a roll structure}
\label{tab:performanceroll}
\renewcommand{\arraystretch}{1.5}
\resizebox{\textwidth}{!}{%
\begin{tabular}{llcccccc}
                             &             & \multicolumn{3}{c}{Low task complexity} & \multicolumn{3}{c}{Medium task complexity} \\ \cmidrule(l{2pt}r{2pt}){3-5}  \cmidrule(l{2pt}r{2pt}){6-8}  
                             &             & \multicolumn{3}{c}{Learning}         & \multicolumn{3}{c}{Learning}             \\ 
Group composition              & Performance & Zero       & Moderate    & High      & Zero        & Moderate     & High        \\ \hline
\multirow{2}{*}{Long-term}   & Mean        & 0.8045    & 0.9251     & 0.9432   & 0.7310     & 0.8896      & 0.8992     \\
                             & Final       & 0.8053    & 0.9542     & 0.9567   & 0.7314     & 0.9200      & 0.9199     \\ \hline
\multirow{2}{*}{Medium-term} & Mean        & 0.8287    & 0.9434     & 0.9475   & 0.7779     & 0.9070      & 0.8981     \\
                             & Final       & 0.8313    & 0.9566     & 0.9568   & 0.7822     & 0.9301      & 0.9210     \\ \hline
\multirow{2}{*}{Short-term}  & Mean        & 0.8278    & 0.9373     & 0.9353   & 0.7782     & 0.8506      & 0.8296     \\
                             & Final       & 0.8292    & 0.9444     & 0.9387   & 0.7822     & 0.8577      & 0.8348     \\ \hline
\end{tabular}%
}
\end{table}

Table \ref{tab:learningroll} shows whether and how a variation in individual learning after promoting group adaptation affects the mean and final performances under the condition of a roll structure. Similarly, Tab. \ref{tab:adaptationroll}  shows whether promoting group adaptation after promoting individual learning has significant effects on the mean and final performances. The general patterns described in Sec. \ref{sec:results} hold for true the roll structure.

\begin{table}[!htbp]
\caption{Significance test for the effects of promoting individual learning for a roll structure}
\label{tab:learningroll}
\renewcommand{\arraystretch}{1.5}
\resizebox{\textwidth}{!}{%
\begin{tabular}{llcccc}
                          &             & \multicolumn{2}{c}{Low task complexity} & \multicolumn{2}{c}{Medium task complexity} \\ \cmidrule(l{2pt}r{2pt}){3-4} \cmidrule(l{2pt}r{2pt}){5-6} 
                             &             & \multicolumn{2}{c}{Learning}         & \multicolumn{2}{c}{Learning}             \\ 
Group composition              & Performance            & Zero to moderate  & Moderate to high & Zero to moderate    & Moderate to high   \\ \hline
\multirow{2}{*}{Long-term}   & Mean & **                & **               & **                  & n.s.               \\
                             & Final & **                & n.s.             & **                  & n.s.               \\ \hline
\multirow{2}{*}{Medium-term} & Mean &  **                 & n.s.             & **                  & n.s.               \\
                             & Final & **                & n.s.             & **                  & n.s.               \\ \hline
\multirow{2}{*}{Short-term}  & Mean  &  **                 & n.s.             & **                  & **                 \\
                             & Final  & **                & n.s.             & **                  & **                 \\ \hline
\end{tabular}%
}
** Indicates significance at the 99\% level\\
n.s. Indicates not significant
\end{table}

\begin{table}[!htbp]
\caption{Significance test for the effects of promoting group adaptation for a roll structure}
\label{tab:adaptationroll}
\renewcommand{\arraystretch}{1.5}
\resizebox{\textwidth}{!}{%
\begin{tabular}{llcccc}
                          &             & \multicolumn{2}{c}{Low task complexity} & \multicolumn{2}{c}{Medium task complexity} \\ \cmidrule(l{2pt}r{2pt}){3-4} \cmidrule(l{2pt}r{2pt}){5-6} 
                             &             & \multicolumn{2}{c}{Group composition}         & \multicolumn{2}{c}{Group composition}             \\ 
Group composition           & Performance            & \Centerstack{Long-term to \\ medium-term}     & \Centerstack{Medium-term to \\ short-term}  & \Centerstack{Long-term to \\ medium-term}    & \Centerstack{Medium-term to \\ short-term}   \\ \hline
\multirow{2}{*}{Zero}     & Mean & **               & n.s.              & **                 & n.s.                \\
                          & Final & **               & n.s.              & **                 & n.s.                \\ \hline
\multirow{2}{*}{Moderate} & Mean & **               & n.s.              & **                 & **                  \\
                          & Final & **               & **                & n.s.               & **                  \\ \hline
\multirow{2}{*}{High}     & Mean & **               & **                & n.s.               & **                  \\
                          & Final & n.s.             & **                & n.s.               & **                  \\ \hline
\end{tabular}%
}
** Indicates significance at the 99\% level\\
n.s. Indicates not significant
\end{table}

\section{Alternative incentive schemes}\label{sec:incentive}
\setcounter{figure}{0} 
\setcounter{table}{0} 

Tables \ref{tab:performancealpha} and \ref{tab:performancebeta} show the mean and final performances for an incentive scheme that favors individualism (i.e., $\alpha=0.75$ and $\beta=0.25$) and collectivism (i.e., $\alpha=0.25$ and $\beta=0.75$), respectively. For decomposed tasks, the performances are very similar to those reported in Tab. \ref{tab:performance}. For interdependent tasks, an incentive scheme that favors individualism decreases group performance. By contrast, an incentive scheme that favors collectivism is associated with higher performances.

\begin{table}[!htbp]
\caption{Mean and final performances for an incentive scheme that favors individualism}
\label{tab:performancealpha}
\renewcommand{\arraystretch}{1.5}
\resizebox{\textwidth}{!}{%
\begin{tabular}{llcccccc}
                             &             & \multicolumn{3}{c}{Decomposed tasks} & \multicolumn{3}{c}{Interdependent tasks} \\ \cmidrule(l{2pt}r{2pt}){3-5}  \cmidrule(l{2pt}r{2pt}){6-8}   
                             &             & \multicolumn{3}{c}{Learning}         & \multicolumn{3}{c}{Learning}             \\  
Group composition              & Performance & Zero       & Moderate    & High      & Zero        & Moderate     & High        \\ \hline
\multirow{2}{*}{Long-term}   & Mean        & 0.8699    & 0.9748     & 0.9883   & 0.7444     & 0.8901      & 0.9014     \\
                             & Final       & 0.8709    & 0.9994     & 1.0000   & 0.7448     & 0.9182      & 0.9163     \\ \hline
\multirow{2}{*}{Medium-term} & Mean        & 0.8706    & 0.9916     & 0.9933   & 0.7892     & 0.9015      & 0.9029     \\
                             & Final       & 0.8717    & 1.0000     & 1.0000   & 0.7945     & 0.9233      & 0.9166     \\ \hline
\multirow{2}{*}{Short-term}  & Mean        & 0.8720    & 0.9946     & 0.9963   & 0.7878     & 0.8869      & 0.8736     \\
                             & Final       & 0.8730    & 1.0000     & 1.0000   & 0.7880     & 0.8934      & 0.8805     \\ \hline
\end{tabular}%
}
\end{table}

\begin{table}[!htbp]
\caption{Mean and final performances for an incentive scheme that favors collectivism}
\label{tab:performancebeta}
\renewcommand{\arraystretch}{1.5}
\resizebox{\textwidth}{!}{%
\begin{tabular}{llcccccc}
                             &             & \multicolumn{3}{c}{Decomposed tasks} & \multicolumn{3}{c}{Interdependent tasks} \\ \cmidrule(l{2pt}r{2pt}){3-5}  \cmidrule(l{2pt}r{2pt}){6-8}  
                             &             & \multicolumn{3}{c}{Learning}         & \multicolumn{3}{c}{Learning}             \\ 
Group composition              & Performance & Zero       & Moderate    & High      & Zero        & Moderate     & High        \\ \hline
\multirow{2}{*}{Long-term}   & Mean        & 0.8722    & 0.9742     & 0.9882   & 0.7502     & 0.9163      & 0.9290     \\
                             & Final       & 0.8732    & 0.9992     & 1.0000   & 0.7507     & 0.9425       & 0.9429     \\ \hline
\multirow{2}{*}{Medium-term} & Mean        & 0.8719    & 0.9914     & 0.9933   & 0.7831     & 0.9286      & 0.9249     \\
                             & Final       & 0.8729    & 1.0000     & 1.0000   & 0.7876     & 0.9435      & 0.9356     \\ \hline
\multirow{2}{*}{Short-term}  & Mean        & 0.8703    & 0.9944     & 0.9963   & 0.7853     & 0.8644      & 0.8444     \\
                             & Final       & 0.8714    & 1.0000     & 1.0000   & 0.7853     & 0.8700      & 0.8470     \\ \hline
\end{tabular}%
}
\end{table}

Tables \ref{tab:learningalpha} and \ref{tab:learningbeta} show whether promoting individual learning after promoting group adaptation has significant effects on the mean and final performances, if the incentive scheme favors individualism and collectivism, respectively. Similarly, Tabs. \ref{tab:adaptationalpha} and \ref{tab:adaptationbeta} show whether promoting group adaptation after promoting individual learning significantly affects task performance, if the incentive scheme favors individualism and collectivism, respectively. The results suggest that the general patterns described in Sec. \ref{sec:positive_effects_learning} and \ref{subsec:group-policy} hold true for alternative configurations of the incentive scheme.

\begin{table}[!htbp]
\caption{Significance test for the effects of promoting individual learning for an incentive scheme that favors individualism}
\label{tab:learningalpha}
\renewcommand{\arraystretch}{1.5}
\resizebox{\textwidth}{!}{%
\begin{tabular}{llcccc}
                             &             & \multicolumn{2}{c}{Decomposed tasks} & \multicolumn{2}{c}{Interdependent tasks} \\ \cmidrule(l{2pt}r{2pt}){3-4} \cmidrule(l{2pt}r{2pt}){5-6} 
                             &             & \multicolumn{2}{c}{Learning}         & \multicolumn{2}{c}{Learning}             \\ 
Group composition              & Performance            & Zero to moderate     & Moderate to high & Zero to moderate    & Moderate to high   \\ \hline
\multirow{2}{*}{Long-term}   & Mean  & **                   & **               & **                  & **                 \\
                             & Final  & **                   & n.s.             & **                  & n.s.               \\ \hline
\multirow{2}{*}{Medium-term} & Mean & ** & n.s.             & **                  & n.s.               \\
                             & Final  & **                   & n.s.             & **                  & n.s.               \\ \hline
\multirow{2}{*}{Short-term}  & Mean  & ** & n.s.             & **                  & **                 \\
                             & Final & **                   & n.s.             & **                  & **                 \\ \hline
\end{tabular}%
}
** Indicates significance at the 99\% level\\
n.s. Indicates not significant
\end{table}

\begin{table}[!htbp]
\caption{Significance test for the effects of promoting group adaptation for an incentive scheme that favors individualism}
\label{tab:adaptationalpha}
\renewcommand{\arraystretch}{1.5}
\resizebox{\textwidth}{!}{%
\begin{tabular}{llcccc}
                          &             & \multicolumn{2}{c}{Decomposed tasks} & \multicolumn{2}{c}{Interdependent tasks} \\ \cmidrule(l{2pt}r{2pt}){3-4} \cmidrule(l{2pt}r{2pt}){5-6} 
                             &             & \multicolumn{2}{c}{Group composition}         & \multicolumn{2}{c}{Group composition}             \\ 
Group composition           & Performance            & \Centerstack{Long-term to \\ medium-term}     & \Centerstack{Medium-term to \\ short-term}  & \Centerstack{Long-term to \\ medium-term}  & \Centerstack{Medium-term to \\ short-term}  \\ \hline
\multirow{2}{*}{Zero}     & Mean  & **               & n.s.              & **                 & n.s.                \\
                          & Final & **               & n.s.              & **                 & n.s.                \\ \hline
\multirow{2}{*}{Moderate} & Mean  & **               & n.s.              & **                 & **                  \\
                          & Final  & **               & **                & n.s.               & **                  \\ \hline
\multirow{2}{*}{High}     & Mean & **               & **                & n.s.               & **                  \\
                          & Final & n.s.             & **                & n.s.               & **                  \\ \hline
\end{tabular}%
}
** Indicates significance at the 99\% level\\
n.s. Indicates not significant
\end{table}

\begin{table}[!htbp]
\caption{Significance test for the effects of promoting individual learning for an incentive scheme that favors collectivism}
\label{tab:learningbeta}
\renewcommand{\arraystretch}{1.5}
\resizebox{\textwidth}{!}{%
\begin{tabular}{llcccc}
                             &             & \multicolumn{2}{c}{Decomposed tasks} & \multicolumn{2}{c}{Interdependent tasks} \\ \cmidrule(l{2pt}r{2pt}){3-4} \cmidrule(l{2pt}r{2pt}){5-6} 
                             &             & \multicolumn{2}{c}{Learning}         & \multicolumn{2}{c}{Learning}             \\ 
Group composition              & Performance            & Zero to moderate  & Moderate to high & Zero to moderate    & Moderate to high   \\ \hline
\multirow{2}{*}{Long-term}   & Mean & **                & **               & **                  & **                 \\
                             & Final & **                & n.s.             & **                  & n.s.               \\ \hline
\multirow{2}{*}{Medium-term} & Mean & **                & n.s.             & **                  & n.s.               \\
                             & Final & **                & n.s.             & **                  & n.s.               \\ \hline
\multirow{2}{*}{Short-term}  & Mean & **                & n.s.             & **                  & **                 \\
                             & Final & **                & n.s.             & **                  & **               \\ \hline 
\end{tabular}%
}
** Indicates significance at the 99\% level\\
n.s. Indicates not significant
\end{table}

\begin{table}[!htbp]
\caption{Significance test for the effects of promoting group adaptation for an incentive scheme that favors collectivism}
\label{tab:adaptationbeta}
\renewcommand{\arraystretch}{1.5}
\resizebox{\textwidth}{!}{%
\begin{tabular}{llcccc}
                          &             & \multicolumn{2}{c}{Decomposed tasks} & \multicolumn{2}{c}{Interdependent tasks} \\ \cmidrule(l{2pt}r{2pt}){3-4} \cmidrule(l{2pt}r{2pt}){5-6} 
                             &             & \multicolumn{2}{c}{Group composition}         & \multicolumn{2}{c}{Group composition}             \\ 
Group composition           & Performance            & \Centerstack{Long-term to \\ medium-term}     & \Centerstack{Medium-term to \\ short-term}  & \Centerstack{Long-term to \\ medium-term}  & \Centerstack{Medium-term to \\ short-term}     \\ \hline
\multirow{2}{*}{Zero}     & Mean & n.s.             & n.s.              & **                 & n.s.                \\
                          & Final & n.s.             & n.s.              & **                 & n.s.                \\ \hline
\multirow{2}{*}{Moderate} & Mean & **               & n.s.              & **                 & **                  \\
                          & Final & n.s.             & n.s.              & n.s.               & **                  \\ \hline
\multirow{2}{*}{High}     & Mean & n.s.             & n.s.              & n.s.               & **                  \\
                          & Final & n.s.             & n.s.              & n.s.               & **                  \\ \hline
\end{tabular}%
}
** Indicates significance at the 99\% level\\
n.s. Indicates not significant
\end{table}
\end{appendices}

\clearpage

\bibliography{sn-bibliography.bib}

\end{document}